\documentclass[preprint]{aastex}
\usepackage{epsf}
\usepackage{graphicx}
\shorttitle{Solar Wind Acceleration and Charge-State Composition in a Global MHD Model}
\shortauthors{Oran et al.}

\begin{document}

\title{A Steady-State Picture of Solar Wind Acceleration and Charge State Composition Derived from a Global Wave-Driven MHD Model}
\author{R. Oran\altaffilmark{1,2},  E. Landi\altaffilmark{1},  B. van der Holst\altaffilmark{1}, S. Lepri\altaffilmark{1}, A. M. V{\'a}squez\altaffilmark{3,4}, F. A. Nuevo\altaffilmark{3,4}, R. Frazin\altaffilmark{1}, W. Manchester\altaffilmark{1}, I. Sokolov\altaffilmark{1} and  T. I. Gombosi\altaffilmark{1}}

\email{oran@umich.edu}

\altaffiltext{1}{Atmospheric, Oceanic and Atmospheric Sciences, University of Michigan, 2455 Hayward, Ann Arbor, MI, 48109, USA}
\altaffiltext{2}{Earth, Atmospheric, and Planetary Sciences, Massachusetts Institute of Technology, 77 Massachusetts Ave., Cambridge, MA, 02139, USA}
\altaffiltext{3}{Instituto de Astronomía y Física del Espacio (IAFE), CONICET-UBA, CC 67 - Suc 28, (1428) Buenos Aires, Argentina}
\altaffiltext{4}{Facultad de Ciencias Exactas y Naturales (FCEN), Universidad de Buenos Aires.}

\begin{abstract}
The higher charge states found in slow ($<$400km s$^{-1}$) solar wind streams compared to fast streams have supported the hypothesis that the slow wind originates in closed coronal loops, and released intermittently through reconnection. Here we examine whether a highly ionized slow wind can also form along steady and open magnetic field lines. We model the steady-state solar atmosphere using AWSoM, a global magnetohydrodynamic model driven by Alfv{\'e}n waves, and apply an ionization code to calculate the charge state evolution along modeled open field lines. This constitutes the first charge states calculation covering all latitudes in a realistic magnetic field. The ratios $O^{+7}/O^{+6}$ and $C^{+6}/C^{+5}$ are compared to in-situ Ulysses observations, and are found to be higher in the slow wind, as observed; however, they are under-predicted in both wind types. The modeled ion fractions of S, Si, and Fe are used to calculate line-of-sight intensities, which are compared to EIS observations above a coronal hole. The agreement is partial, and suggests that all ionization rates are under-predicted. Assuming the presence of suprathermal electrons improved the agreement with both EIS and Ulysses observations; importantly, the trend of higher ionization in the slow wind was maintained. The results suggest there can be a sub-class of slow wind that is steady and highly ionized. Further analysis shows it originates from coronal hole boundaries (CHB), where the modeled electron density and temperature are higher than inside the hole, leading to faster ionization. This property of CHBs is global, and observationally supported by EUV tomography.
 \end{abstract}


\section{Introduction}
The formation of the solar wind and its acceleration through interplanetary space pose some of the central outstanding problems in solar physics. These include identifying the processes by which the solar wind is formed and accelerated, and explaining how these processes produce the observed three-dimensional, time-dependent distributions of plasma properties and composition. 
The solar wind has been measured and analyzed extensively over the last few decades, and considerable amounts of data have been gathered. This has led to the identification of distinctly different solar wind flows, commonly classified as the fast ($\sim$ 700 km s$^{-1}$) or slow ($\sim$ 300 - 400 km s$^{-1}$) solar wind \citep[see e.g., ][]{{McComas2003}}. While it is generally accepted that the fast wind originates from coronal holes (CH), the markedly different chemical composition and temporal variability of the slow wind has led to an on-going and vigorous debate regarding its source region and formation mechanism\citep{kohl2006,suess2009,abbo2010,Antiochos2011,antonucci2011,Antiochos2012}.\par 
The abundances of heavy elements in the solar atmosphere and their ionization state have played a central role in testing theories of solar wind formation. The abundances of elements heavier than helium, relative to that of hydrogen, are lower than 0.001 everywhere in both the solar wind and solar corona \citep[e.g.][]{feldman1992, Asplund2009, Caffau2011}, and therefore their contribution to the large-scale dynamics is negligible. However, their response to the local state of the plasma in which they are embedded makes them useful tracers of the conditions in different regions. Indeed, both their relative abundances and their ionization status vary when observed in different regions of the corona and the wind.\par
The abundances of certain elements are modified in the corona relative to their photospheric values according to their First Ionization Potential (FIP)\citep[c.f.][and references therein]{Feldman2000, Feldman2003}. The ratio of coronal to photospheric abundances is called the FIP bias. Closed-field structures such as helmet streamers and active regions exhibit a FIP bias between 2 and 4 for low-FIP ( $<$10 eV) elements, while CHs do not \citep{Feldman2003}. To date, there is still no clear and conclusive picture that explains the observed FIP bias in the corona, but several promising theories are being developed (see \citet{Laming2009, laming2012}  for a review of this active research area).\par 
In contrast to the FIP bias, the basic mechanisms controlling heavy element ionization are well understood. As the ions propagate away from the Sun, they undergo ionization and recombination due to collisions with free electrons. The collision rate depends on the electron density, while the ionization and recombination rate coefficients can be derived from atomic physics, provided the energy of the electrons is known. Due to the decrease of electron density with distance from the Sun, at a certain distance the plasma becomes collisionless and ionization and recombination processes effectively stop. At this point the charge state distribution of the element is said to ``freeze-in'', which usually occurs at distances between 1.5 to ~4 $R_\odot$, depending on the ion considered \citep{Hundhausen1968}. The charge state distribution, which is routinely analyzed by in-situ measurements in the heliosphere, therefore contains information about the wind evolution very close to the Sun.\par 

In this paper, we examine whether the scenario in which the wind is heated and accelerated by Alfv{\'e}n waves can explain the observed charge state distributions, both in the solar corona and in the fast and slow solar wind. For this purpose, we use a global magnetohydrodynamic (MHD) computational model driven by Alfv{\'e}n waves to predict the plasma flow properties and magnetic field from the transition region to 2AU.  We then calculate the charge state evolution of heavy elements as they flow along modeled open field lines and undergo ionization and recombination. In order to study both slow and fast wind flows, we perform these calculations at all latitudes. As we describe in more detail below, elemental abundances and dynamic processes are not included in the simulations. Nonetheless, comparing the modeled charge state distributions to available in-situ and remote observations will allow us to gain further insight into how well the MHD model describes the wind evolution, and to extend our current understanding of how and where the slow wind is formed.

\subsection{Theoretical Models of Solar Wind Formation}
A wide range of theoretical models relate the distribution of fast and slow wind speeds to the steady state magnetic field geometry and the expansion of magnetic flux tubes \citep{Suess1979,Kovalenko1981,Withbroe1988,Wang1990, roussev2003, cranmer2005, suzuki2006, cranmer2007, cohen2007, vanderholst2010}. In this picture, both the fast and slow wind flows along static open field lines, and the slow wind originates from the boundary region between CHs and closed field lines, where the expansion is largest. However, static expansion models by themselves cannot explain the different chemical composition of the slow wind and fast wind: the fast wind exhibits elemental abundances characteristic of the photosphere and CHs \citep{vonSteiger2001, Zurbuchen1999, Zurbuchen2002}, while the slow wind exhibits FIP-biased abundances similar to that of closed coronal loops \citep{Feldman2003}. In addition, the charge states measured in the fast wind are compatible with a coronal electron temperature of $\sim$1.0MK, similar to that occurring in CHs \citep[e.g.][]{Gloeckler2003, Zurbuchen2007}, while the charge states in the slow wind are generally higher, and are compatible with higher coronal electron temperatures, as found in closed-field regions \citep[e.g. ][]{Gloeckler2003, Zurbuchen2002}. \par  
The correspondence between slow wind composition and the properties of coronal loops has led to the hypothesis that the slow wind plasma originates in the hotter and denser closed field region in the corona. These models suggest that the plasma is dynamically and intermittently released into space due to reconnection between open and closed field lines, although the details and the location of the reconnection process vary (e.g. the Interchange Reconnection Model, \citep{Fisk1998, Fisk2003, FiskZhao2009}; the Streamer-Top Model, \citep{Wang2000}; the S-web Model, \citep{Antiochos2007,Antiochos2011,Antiochos2012}). Dynamic release models can also potentially explain the different levels of fluctuations observed in the fast and slow wind. The flow properties of the fast wind are relatively steady \citep[e.g. ][]{McComas2008}, while those measured in the slow wind are highly variable \citep{Schwenn1990,Gosling1997,McComas2000}. Similarly, the chemical composition of the fast wind is relatively steady \citep{Geiss1995,
vonSteiger1995, Zurbuchen2007}, while that of the slow wind is highly variable \citep{Zurbuchen2006, Zurbuchen2007}. Dynamic release models offer a natural explanation for this variability, since they imply that the slow wind is formed in a series of discrete and localized release events. Dynamic release models are limited by the fact that the localized and time-dependent nature of the release process make it difficult to produce global simulations with a realistic magnetic field.\par

Another class of solar wind acceleration models are wave-driven models. \citet{alazraki1971} and \citet{belcher1971} have suggested that low frequency Alfv{\'e}n waves emanating from the chromosphere can accelerate the wind due to gradients in the wave pressure, and heat the corona through wave dissipation. The steep density gradient in the transition region will cause a significant amount of the wave energy to be reflected. However, radiative-MHD simulations by  \citet{depontieu2007} have shown that between 3\% to 15\% of the observed chromospheric wave energy will be transmitted into the corona, with a resulting energy flux that is sufficient to sustain the corona and solar wind. Indeed, Alfv{\'e}nic perturbations are ubiquitous in the solar environment, and have been observed in the photosphere, chromosphere, in coronal structures, and in the solar wind at Earth's orbit \citep[c.f.][]{banerjee2011,mcintosh2011}.\par 
Alfv{\'e}n waves were incorporated into several magnetohydrodynamic (MHD) models of the solar atmosphere in an attempt to explain the observed properties of the solar wind and corona \citep[e.g. ][ to name a few]{Usmanov2000,cranmer2005,cranmer2007, vanderholst2010, evans2012,sokolov2013,oran2013,vanderholst2014, Lionello2014a, Lionello2014b}. These models were able to describe the large-scale features of the corona and the wind, but for the large part did not explicitly address the wind's composition \citep[except][which will be discussed below]{cranmer2007} or the temporal variability .

\subsection{The Goal and Context of this Paper} 
The goal of this work is twofold: first, we wish to examine whether a solar wind model in which the wind is accelerated by Alfv{\'e}n waves can explain the charge state distributions observed in both the corona and the wind. Second, we address the question of whether a solar wind which originates solely from CHs and propagates along static open magnetic field lines can lead to the formation of higher charge states in slow flows compared to fast flows, without invoking dynamic release from the closed field region.\par
We use the Alfv{\'e}n Wave Solar Model \citep[AWSoM, ][]{sokolov2013, oran2013, vanderholst2014}, which extends from the top of the transition region up to 2AU. The model solves the two-temperature (electrons and protons) MHD equations coupled to wave transport equations of parallel and anti-parallel Alfv{\'e}n waves. Wave propagation and dissipation are treated self-consistently in both open and closed field regions, as described in \citep{sokolov2013}. \citet{oran2013} showed that for a solar minimum configuration, the model can reproduce remote observations of the lower corona simultaneously with the large scale distribution of wind speeds observed by Ulysses at 1-2 AU.\par 
 We take advantage of the steady-state simulation of the solar atmosphere previously presented and validated in \citet{oran2013} as a basis for modeling charge state evolution and comparing the results to in-situ and remote observations. The simulation was constrained by a synoptic map of the photospheric magnetic field observed during Carrington Rotation (CR) 2063, which took place during solar minimum. The electron density, temperature and speed from the MHD simulation are used as input to a charge state evolution model \citep[Michigan Ionization Code (MIC), ][]{landi2012b} which calculates the ionization status of an element at any point along the wind trajectory. We calculate the evolution of C, O, S, Si, and Fe charge states, in order to compare the results to as many available observations as possible, both in the corona and in the wind.\par
The steady-state simulations presented here cannot describe dynamic release of material from closed field structures, and we do not aim to examine how well these models explain the observations. In fact, in a static magnetic field both the slow and fast wind must originate from coronal holes and flow solely along open field lines. In this sense, the simulation presented here can be grouped with the expansion models. \citet{Antiochos2012} argued that expansion models cannot give a complete picture of solar wind formation, as they cannot explain the different composition and the large temporal fluctuations observed in the slow wind. A static wind model indeed cannot explain the different elemental abundances or the fluctuations; however, the question still remains, whether a wind flowing along static open field lines can posses a large scale variation in charge states, solely because ions flowing along different trajectories will encounter different plasma conditions, and will therefore have different ionization histories. \par 
 \citet{cranmer2007} calculated the charge state evolution of O ions in a steady solar wind model driven by turbulent waves, and found the resulting ion fraction to be in qualitative agreement with Ulysses observations. The \citet{cranmer2007} model is based on a prescribed axially symmetric magnetic field topology that is not derived self-consistently with the plasma and wave field. This limits the analysis to idealized flux tube geometries, and cannot include more complex structures. \citet{jin2012} calculated the frozen-in charge state distributions using a 3D MHD model with a realistic and self-consistent magnetic field. The calculation was performed over a few representative field lines and was not aimed to address the variation between fast and slow wind streams. Here we present the first calculation of charge state distributions covering all heliographic latitudes, in a realistic, fully three-dimensional and self-consistent magnetic field configuration. This allows us to examine how the modeled frozen-in distributions vary with terminal wind speed, study the evolution below the freeze-in height, and compare the results with observations taken concurrently with the photopsheric magnetogram driving the simulation.\par
The modeled frozen-in distributions for O and C will be directly compared to in-situ measurements performed by the Solar Wind Ion Composition Spectrometer \citep[SWICS,][]{Gloeckler1992} on board Ulysses taken during its third polar scan at a distance of 1-2AU. In the lower corona, on the other hand, information about the ionization state can only be gained from the observed emission associated with the different ions. We derive synthetic line intensities for S, Si and Fe ions from the model and compare them to remote observations made by the EUV Imaging Spectrometer \citep[EIS,][]{culhane2007} on board Hinode. Several spectral lines corresponding to different ionization stages are used, which allows us to examine the modeled ionization in detail. The simultaneous comparison to both remote and in-situ observations allows us to test the predicted charge states as both ends of the wind trajectory \citep{landi2012a}. This diagnostic approach was used by \citet{Landi2014} to test predictions of three theoretical models, including the AWSoM model, by applying the MIC code to a  field line stretching along the center of a polar CH in an ideal dipole field. The strength of the 3D nature of the AWSoM-MIC simulations presented here, is that we can calculate the charge states and their emission at every point along the line of sight, allowing us to produce synthetic emission profiles without the need to make simplifying assumptions about the spatial variation of these properties. This makes for a more rigorous model-data comparison.\par 
\par
Finally, we note that this work does not address the variation of elemental abundances observed in the fast and slow wind. Describing the formation of the FIP bias in an MHD model will require: 1. a multi-fluid description to describe the evolution of each element, and 2. the inclusion of an elemental fractionation mechanism responsible for the FIP bias, which as of yet has not been conclusively identified, and 3. a time-dependent description of coronal morphology. The last requirement stems from the fact that the FIP bias is known to vary with the age of a coronal loop, i.e. the time elapsed since its emergence from the chromosphere \citep[e.g.][]{Feldman2003}. A steady-state model driven by a synoptic magnetogram of the photospheric field cannot account for temporal changes. In addition, the FIP bias is largely active in lower and cooler regions of the solar atmosphere, and proper modeling of its creation would require a realistic model of the chromosphere, which is not included in the present AWSoM model. For these reasons, we defer the question of elemental abundances to future work, and only address the charge state composition.\par 
This paper is organized as follows. The theory of charge state evolution and the MIC code are described in Section \ref{s:MIC}. The AWSoM model and the steady-state simulation used in this paper are presented in Section \ref{S:AWSoM}. We discuss how the AWSoM simulation results were used to drive the ionization code in Section \ref{S:fieldlines}. The method of creating synthetic emission from the AWSoM-MIC results is described in Section \ref{s:synth}. The in-situ and remote observations used in this work are presented in Section \ref{s:observations}. We present the model results and their comparison to the observations in Section \ref{s:results}. Section \ref{s:ocb} discusses the main result of this paper, i.e. the formation of higher charge states in the modeled steady slow wind.  We describe the different source regions of these wind streams, and discuss how the plasma properties close to the Sun explain the increased ionization. We show that the main component of this steady slow wind, which come from the boundaries of CHs, is highly ionized due to increased electron density and temperature compared to deeper inside the holes, and present observational evidence for this enhancement from an EUV tomographic reconstructions of the lower corona. Section \ref{s:discussion_ch4} summarizes the results and discusses their possible interpretations and implications to understanding solar wind formation.

\section{Charge State Evolution Model}\label{s:MIC}
\subsection{Evolution Along Field Lines}
As heavy ions are accelerated away from the Sun, they undergo ionization and recombination due to collisions with the electrons, at rates that depend on the local electron density, $N_e$, and temperature, $T_e$. The speed of the ions determines how much time they spend at a given location; if the speed is sufficiently high, the ions will not reach local ionization equilibrium. In this case the population of each charge state can only be determined by taking into account the flow properties along the entire trajectory. The rate of change (in the rest frame) of the population of element $y$ at charge state $m$ is given by the following equation \citep{Hundhausen1968}:

\begin{eqnarray}\label{eq:mic}
\frac{\partial  N_y y_m}{\partial t}+\nabla\cdot(y_m N_y{\bf u})=N_eN_y
[y_{m-1}C_{m-1}(T_e)+y_{m+1}R_{m+1}(T_e)-y_mC_m(T_e)-y_mR_m(T_e)],\nonumber \\
\sum_m y_m = 1,  
\end{eqnarray}
where $N_y$ is the total number density of element $y$, $y_m$ is the fraction of element $y$ in charge state $m$, $R_m$ and $C_m$ are recombination and ionization rate coefficients, respectively, and ${\bf u}$ is the 
ion velocity. The first two terms on the right hand side describe the creation of ions with charge state $m$ due to ionization from a lower charge state and recombination from a higher charge state, while the last two terms describe losses due to ionization and recombination of ions with charge $m$ into higher and lower charge states, respectively. Ionization and recombination are assumed to be due to binary reactions between ions and electrons, namely direct collisional ionization, excitation-autoionization, radiative recombination, and dielectronic recombination. Three-body
recombination (as well as photoionization) are negligible in the solar atmosphere \citep{Hundhausen1968}. Thus in Eq. (\ref{eq:mic}) the number of reactions occurring per unit volume per unit time is proportional to the product of the concentrations of the reacting 
particles, $N_eN_y y_m$. The recombination and ionization rate coefficient depend on the electron energy and are calculated using the CHIANTI 7.1 Atomic Database \citep{Dere1997,Landi2013}. The rate coefficients in CHIANTI are largely based on the ionization rates compiled by \citet{Dere2007} and the recombination rates reviewed by \citet{Dere2009}. \par 
Eq. (\ref{eq:mic}) constitutes a system of continuity equations of the number density of each
charge state, which are coupled through the ionization and recombination source terms. Taking the sum of all the equations for each element, we obtain a 
continuity for the total elemental number density $N_y$:
\begin{equation}\label{eq:ny}
\frac{\partial N_y}{\partial t}+ \nabla\cdot(N_y{\bf u})=0.
\end{equation}
On dividing each of the continuity equations in Eq. (\ref{eq:mic})  by $N_y$ and subtracting Eq. (\ref{eq:ny}), we obtain the following system of equations:
\begin{eqnarray}\label{eq:mic2}
({\bf u}\nabla\cdot)y_m=u_\|\frac{dy_m}{ds}=N_e
[y_{m-1}C_{m-1}(T_e)+y_{m+1}R_{m+1}(T_e)-y_mC_m(T_e)-y_mR_m(T_e)],\nonumber \\
\sum_m{y_m}=1,
\end{eqnarray}
where $u_\|$ as the speed parallel to the flow line and $ds$ is the path length. This system of equations is solved numerically by the MIC code using a fourth-order Runge-Kutta method with an adaptive step size which limits the change in any charge state fraction to a maximum of $10\%$. The boundary conditions for $y_m$ at the base of the flow line are derived assuming ionization equilibrium. \par
The MIC model requires information about the electron density and temperature as well as the wind speed in order to solve Eq. (\ref{eq:mic2}). In this work we extracted these from the MHD solution given by the AWSoM model. In the MHD approximation,  plasma flows parallel to magnetic field lines in the rest frame of the plasma, which in our case is the frame co-rotating with the Sun. We extract the needed quantities along open magnetic field lines, and $u_\|$ is taken with respect to the co-rotating frame. Since we are interested in the large-scale steady-state solution, the wind properties at any point are constant in time. The AWSoM model equations do not describe the ion motion, and it is therefore assumed that the ions move with the same speed as the plasma. This assumption does not strictly hold at all locations in the solar atmosphere, and future work may take differential ion speeds into account.
\subsection{Role of Supra-Thermal Electrons}\label{s:supertheory}
Supra-thermal electrons can have a considerable effect on charge state evolution, as their energy will modify the ionization  rate coefficients. As of yet, there is no direct observational evidence of their presence in the lower corona, and the subject is still under debate \citep[see][for a review]{cranmer2009}. However, a supra-thermal population can potentially reconcile the discrepancy between the observed charge states and coronal temperatures. Several studies used the observed frozen-in charge states in the fast wind in order to put constraints on the electron temperature low in coronal holes \citep[see e.g.][]{Geiss1995, Ko1997}.  When a purely Maxwellian electron population was assumed, the coronal temperatures that can explain the in-situ observations were about 50$\%$ higher than those derived from spectral observations below the freeze-in height. \citet{Esser2000} showed that this discrepancy can be resolved if an additional small population of supra-thermal electrons is present. Differential ion speeds may have a similar effect on the frozen-in charge states \citep{Ko1998, Esser2001}, but this mechanism is beyond the scope of the present work. \citet{Laming2007} showed that supra-thermal electrons can be created due to parallel heating by lower hybrid wave damping, giving rise to a kappa distribution function for the electrons, which can explain the observed charge states. \citet{Feldman2007} estimated the energy content of supra-thermal electrons in an active region, and found that less than $5\%$ of the electron population can have energies above 0.91keV and less than $2\%$  can have energies above 1.34keV in active regions. \par 
Following these previous efforts, in this work we consider the charge state evolution due to a single temperature plasma as well as a plasma with an additional hotter electron population, in order to evaluate their contribution.
We assume that $2\%$ of the electrons belong to a second Maxwellian distribution at $3MK \approx 0.25keV$. These parameters were chosen empirically as we describe in Section \ref{s:results}. Ideally, a full parametric study of these values should be performed, guided by observations. Such a study is beyond the scope of this work. Nonetheless, incorporating the supra-thermal electrons in the simulation serves as a proof of concept, to determine whether they can, at the same time:\\ 
\indent 1. affect the predicted charge state composition and improve the agreement with in-situ observations; and \\
\indent 2. produce observable signatures in coronal emission (to our knowledge, such signatures were not found to date), and that their effect on the emission is consistent with observed spectra.\par
In order to accomplish this, we need to apply two sets of ionization rate coefficients when solving Eq. (\ref{eq:mic2}): one in which only the thermal electron population is taken into account, and another where both the thermal and supra-thermal populations are considered. Supra-thermal electrons will also impact the emissivity of the plasma, and therefore we take them into account when calculating synthetic emission from the model, as we describe in Section \ref{s:synth}.
\section{The AWSoM Model Description}\label{S:AWSoM}
The AWSoM model is a three-dimensional computational model of the solar environment, extending from the transition region into inter-planetary space. It solves the extended-MHD equations (with separate electron and proton temperatures) coupled to wave transport equations for low-frequency Alfv{\'e}n waves, propagating parallel or anti-parallel to the magnetic field. The coupled equations allow for a self-consistent description of coronal heating and wind acceleration, where wave dissipation heats the plasma and wave-pressure gradients accelerate it. Wave dissipation is the only heating mechanism, and the dissipated energy is partitioned between the protons and electrons. The separate electron and proton temperatures enable us to include non-ideal MHD processes: field-aligned electron heat conduction, radiative cooling, and thermal coupling between the electrons and protons.\par 
A detailed description of the model and its development was presented in \citet{sokolov2013, oran2013, vanderholst2014}. The AWSoM simulation used in this work is described in detail in \citet{oran2013}. The wave dissipation is assumed to be a result of fully-developed turbulent cascade \citep{matthaeus1999} due to counter propagating waves in closed field regions and wave reflections in open field regions. Wave reflections, which are in general frequency dependent, are not described explicitly \citep[as was done, for example, in][]{ cranmer2005, cranmer2007}. Rather, the model adopts the approach proposed by \citet{Hollweg1986}, in which a Kolmogorov-type dissipation rate is assumed. The Kolmogorov approach, originally developed for open magnetic flux tubes, was generalized to both open and closed field lines in \citet{sokolov2013}. The dissipation mechanism was analyzed in detail in \citet{sokolov2013,oran2013}, and its predictions of the wave amplitude in the corona and solar wind were shown to be consistent with observation both in the solar wind \citep{oran2013}, and in the lower corona \citep{oran2014} during solar minimum.  \citet{jin2013} simulated a more complex magnetic topology which took place during the ascending phase of the solar cycle. They successfully simulated the propagation and evolution of a coronal mass ejection, whose modeled evolution was validated against white-light observations of the outer corona.\par 
In this work we use an AWSoM simulation for CR2063, which took place between November 4 and December 1 in 2007. The boundary conditions for the radial magnetic field are driven from a line-of-sight synoptic magnetogram obtained for that period by the Michelson-Doppler Interferometer (MDI) instrument on board the Solar and Heliospheric Observatory (SOHO) spacecraft \citep{Scherrer1995}. The simulation set-up, input parameters and comparison to remote and in-situ observations are described in detail in \citet{oran2013}. \par 
\section{Coordinated Observations and Field Line Selection}\label{S:fieldlines}
We take advantage of high resolution observations performed by the EIS instrument on board Hinode taken during CR2063, on 2007 November 16, at 11:47:57UT, observing the north polar CH. This particular set of EIS observations was chosen since it includes bright and isolated emission lines from several charge states of Fe, two charge states of Si and one charge state of S. In the same period, Ulysses was performing its third and last polar scan, covering almost all latitudes in a period of a little over a year.\par 
Modeling the charge state evolution for all ions in the entire 3D domain is computationally expensive, and therefore we only solve the charge state evolution along selected field lines, depending on the specific need:\par
1.  For comparison with \textit{remote observations}, we chose the field lines that intersect the EIS line of sight. Field lines at 1 degree spacings in the northern hemisphere were extracted; although they lie in the same meridional plane at altitudes covered by the EIS slit above the north polar CH, they reach slightly different longitudes at their foot points, due to the complex magnetic topology. \par
2.  For comparison with Ulysses observations, the MIC solution is obtained for field lines that reach the same meridional plane at 1.8AU, at all latitudes at 1 degree spacings. Since AWSoM is driven by a synoptic magnetogram, changes in the solar magnetic field during the year-long Ulysses polar scan are not simulated. The comparison should be regarded as a qualitative examination of how well the model reproduces the large-scale structure of the frozen-in charge states during solar minimum. In this case tracking the solution along the field lines reaching the exact Ulysses trajectory  is not needed, and it suffices to cover all latitudes.
\par
\begin{figure}[hp]
\epsscale{0.9}
\plotone{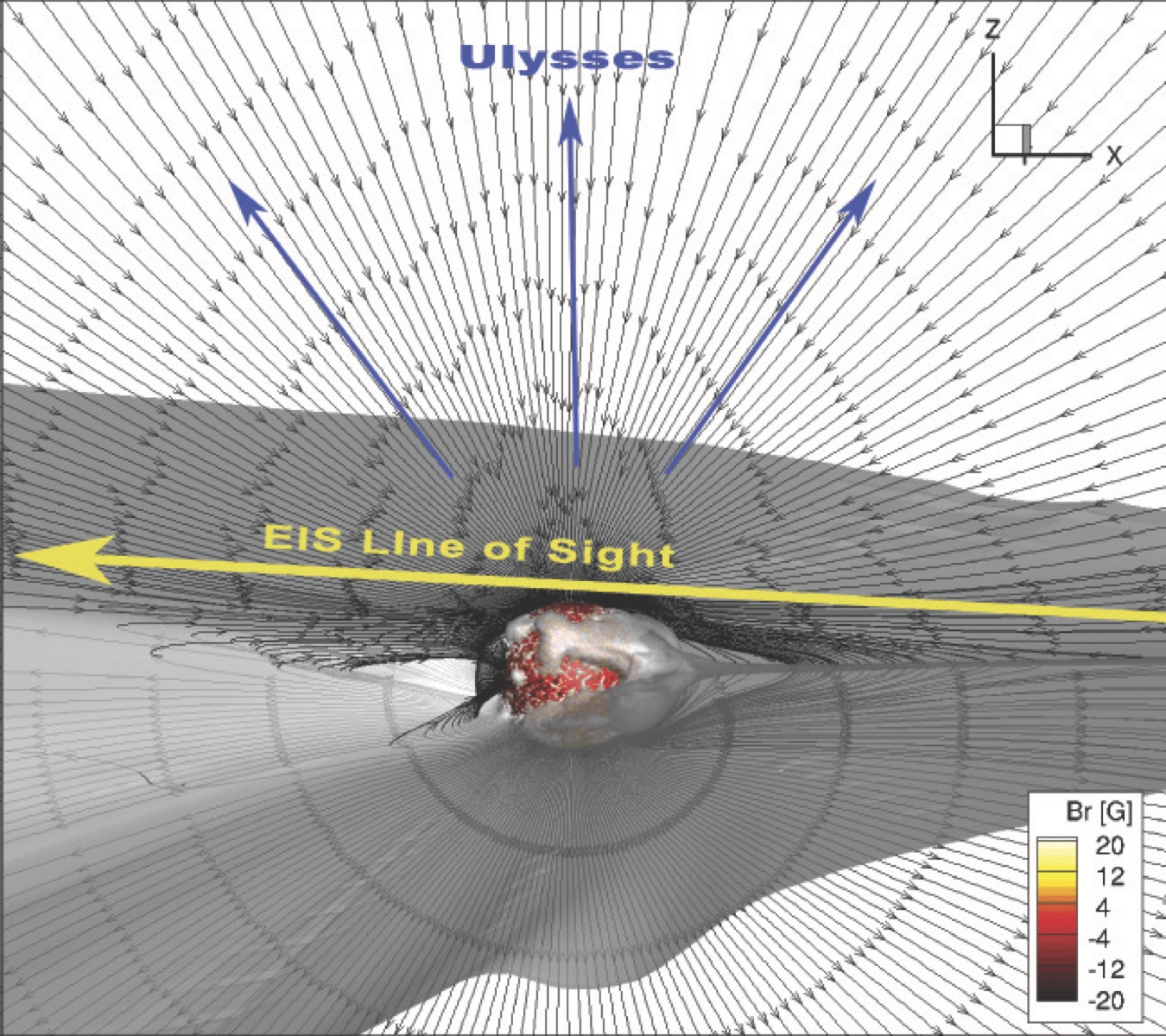} 
\caption{\small \sl Geometry used for comparing model results with Ulysses and EIS coordinated observations. Black stream lines show the magnetic field lines extracted from the AWSoM simulation for CR2063. Wind parameters along the open field lines were used as input to MIC. Labeled arrows mark the direction of the EIS line of sight and the general direction of Ulysses during its polar scan. The solar surface is colored by the radial magnetic field obtained from a synoptic GONG magnetogram. The gray surface represents the heliospheric current sheet, where the radial magnetic field is zero. \label{F:geometry}}
\end{figure}

The geometry is shown schematically in Figure \ref{F:geometry}. The black curves are magnetic field lines, while the solar surface is colored by the radial magnetic field and the gray surface represents the location of the current sheet. The direction of the EIS line of sight is shown by the yellow arrow. The blue arrows represent the general direction of Ulysses polar pass, although the details of the trajectory itself are not represented in this figure. Note that only the open field lines were used to obtain a solution from MIC, and closed field lines are shown here for clarity.  \par

\section{Calculating Synthetic Line of Sight Emission due to Non-Equilibrium Ion Fractions}\label{s:synth}
The emission of a volume of plasma at a given spectral line due to an electronic transition from an upper level $j$ to a lower level $i$ depends on the contribution function, $G_{ji}(N_e,T_e)$, defined as:
\begin{equation}\label{eq:GNT}
G_{ji}(N_e,T_e)= A_{ji}\frac{N_j(X^{+m})}{N(X^{+m})}\frac{N(X^{+m})}{N(X)}\frac{N(X)}{N(H)}\frac{N(H)}{N_e}\frac{1}{N_e},
\end{equation}
where $G_{ji}$ is measured in units of photons cm$^{3}$ s$^{-1}$. $X^{+m}$ denotes the ion of the element $X$ at ionization state $+m$. The separate terms are defined as:\par 
1. $ N_j(X^{+m}) / N(X^{+m})$ is the relative level population of $X^{+m}$ ions at level $j$, and depends on the electron density and temperature ;\par 
2. $N(X^{+m}) / N(X)$ is the abundance of the ion $X^{+m}$ relative to the abundance of the element $X$;\par  
3. $N(X) / N(H)$ is the abundance of the element X relative
to hydrogen ;\par 
4. $N(H) /N_e$ is the hydrogen abundance relative to the electron density ($\approx$0.83 for a fully ionized plasma); and\par 
5. $A_{ji}$ is the Einstein coefficient for spontaneous emission for the transition $j\rightarrow i$.\par 
The contribution function in any computational volume element can be calculated by combining the modeled electron density and temperature with the CHIANTI atomic data base. For the elemental abundances (term 3) we used the photospheric elemental abundances from \citet{Caffau2011}. The relative ion abundances of the same element (term 2) can be derived by assuming ionization equilibrium at the local plasma conditions. Depending on their speed, the ions may not have sufficient time to reach ionization equilibrium. The AWSoM-MIC simulations allow us to directly predict the ratio $N(X^{+m}) / N(X)$ (which we hereafter refer to as ion fraction)  from the charge state evolution, without assuming ionization equilibrium. \par 
In this work we consider charge state evolution in both a plasma with single-temperature electron population, and in a plasma containing an additional population of supra-thermal electrons (see Section \ref{s:supertheory}). In the latter case, the higher energies of the supra-thermal electrons will result in different ionization and recombination rate coefficients and ultimately in different ion fractions in term 2. The level population, $N_j(X^{+m})/N(X^{+m})$, appearing in term 1, will also be affected by the presence of supra-thermal electrons, as these will change the collisional excitation/de-excitation rates. The modified rates can also be obtained from CHIANTI.\par 
Once the contribution function is calculated at every point along the line-of-sight, the total observed flux in the optically thin limit is given by the integral:
\begin{equation}\label{eq:ftot}
F_{tot} = \int \frac{1}{4\pi d^2} G_{ji}(N_e,T_e)N_e^2 dV ,
\end{equation}
where $d$ is the distance of the instrument from the emitting volume $dV$.  $F_{tot}$ is measured in units of photons cm$^{-2}$ s$^{-1}$. This volume integral can be replaced by a line integral by observing that $dV=Adl$, where $A$ is the area observed by the instrument and $dl$ is the path length along the line of sight (LOS). The electron density, electron temperature and contribution function predicted by AWSoM-MIC are interpolated from the field lines intersecting the LOS into a uniformly spaced set of points along each observed LOS. This procedure ensures that the integration is second-order accurate.
\section{Observations}\label{s:observations}
\subsection{Ulysses in-situ Charge States}
We use the charge state measurements obtained by the SWICS instrument on board Ulysses between 15-Feb-2007 and 15-Jan-2008. This period overlaps the time at which the synoptic magnetogram for CR2063 and the remote EIS observations were obtained. The start and end dates were chosen so that the widest range of latitudes is included in the data set. The charge states ratios of O$^{7+}$/O$^{6+}$ and C$^{6+}$/C$^{5+}$ and the average charge state of Fe, $<Q>_{Fe}$, are publicly available through ESA's Ulysses data system, and their calculation from the raw measurements is described in \citet{vonSteiger2000}. The statistical accuracy of the measurements is estimated to be 10 - 25\% (Ulysses/SWICS Heavy Ion Composition Data: User's Recipe, by T. Zurbuchen and R. von Steiger, 2011).\par 
The oxygen and carbon charge state ratios are sensitive to the electron temperature in the inner corona (up to the freeze-in height of 1.5-2R$_\odot$), and they are often used to distinguish between different solar wind types and to study their source regions \citep[e.g.,][]{Zurbuchen2002, Zhao2009}.
The charge state of Fe have a freeze-in height of $\sim 4$R$_\odot$ and were used to study the wind evolution in the outer corona  \citep[e.g.,][]{Lepri2001, Lepri2004,Gruesbeck2011}. 
However, the magnitude of $<Q>_{Fe}$ does not change by much when measured in the fast and slow wind \citep{Lepri2001}, and its behavior in the two wind types only differs in the level of temporal fluctuations. We therefore focus on O$^{7+}$/O$^{6+}$ and C$^{6+}$/C$^{5+}$ to study how the modeled charge states vary with wind speed.
\subsection{Emission from the Lower Corona}
We use the spectral observations made by the EIS instrument on 16 November 2007. During this 
time, the EIS 2"$\times$512" slit was oriented 
along the North-South direction and was pointed at 7 adjacent position along the solar E-W direction to cover a total field of view of 14"x512" whose center was located at (0",866"). The field of view extended from 0.61~R$_\odot$ from the Sun center inside the disk and up to a height of 1.15~R$_\odot$ above the limb in the north CH. At each location of the raster, the spectral range covered was $171-211$\AA~ and $245-291$\AA~ (with a spectral pixel size of  0.022\AA~ per pixel) and the exposure time was 300s. From the available spectral range, we chose a set of bright and isolated spectral lines (listed in Table \ref{T:linesEIS}), which includes as wide a range of charge states belonging to the same element as possible. More details on these observations can be found in \citet{Hahn2010}.\par 

\begin{table}[ht]
\centering
\begin{tabular}{| l | c | c | c |  }
\hline \hline
Ion Name & Wavelength & $F_{scatt}$ & R$_{max}$    \\
		   & [\AA] & [erg cm$^{-2}$ s$^{-1}$ sr$^{-1}$] &  [R$_\odot$]\\
\hline\hline
Fe VIII &  185.213 & 29.35 & 1.115    \\
\hline
Fe IX & 188.497 & 22.36 & 1.136    \\
\hline
Fe IX & 197.862 & 9.51 & 1.136   \\
\hline 
Fe X & 184.537  &78.01 &  1.136  \\
\hline 
Fe XI & 188.217 & 101.17 & 1.125   \\
\hline
Fe XI & 188.299 & 78.06 &  1.125 \\
\hline 
Fe XII & 195.119 & 121.76 & 1.106  \\
\hline 
Si VII & 275.361 & 14.79 & 1.136   \\
\hline 
Si X & 261.057 & 15.66 & 1.136   \\
\hline 
S X & 264.231 &15.68 & 1.115  \\
\hline \hline 
\end{tabular}
\caption{\small \sl Selected EIS emission lines. $F_{scatt}$ indicates the instrument-scattered light flux and R$_{\rm{max}}$ is the highest altitude at which the scattered flux is less than 20\% of the observed flux (see Section \ref{S:stray4}).}
\label{T:linesEIS}
\end{table}
\subsubsection{Data Reduction and Selection}
The data were reduced using the standard EIS software made available by
the EIS team through the SolarSoft IDL package \citep{freeland1998}. Each original frame was flat-fielded, the dark current and CCD bias were subtracted, the cosmic ray hits were removed, and the defective pixels were flagged. Residual wavelength-dependent offsets and the tilt of the detectors were also removed. Data were calibrated in wavelength and intensity; the most recent EIS intensity calibration from Warren et al. 2014 was applied. This updated intensity calibration improves the calibration of the long-wavelength channel (246-292 A) and also allows to account for the degradation occurred during the EIS mission. The accuracy of the calibration is $\approx$25\%.\par 
In order to increase the signal-to-noise ratio, the data were averaged
along the E-W direction and re-binned along the slit direction (N-S) in bins of 0.01~R$_\odot$.\par 
Only 14 bins extending from 1.025R$_\odot$ to 1.155$_\odot$ above the limb were used for comparison with the model. Pixels between 1.00 - 1.025R$_\odot$ were excluded since they might be affected by limb brightening and spicule material \citep{Hahn2010}. The portion of the slit pointed inside the solar disk was only used for evaluating the instrument-scattered light, as we describe in Section \ref{S:stray4}. \par Spectral line profiles were fitted with a Gaussian curve removing a linear background. At a certain height above the limb the line emission becomes too weak, and a clear Gaussian cannot be discerned; these measurement are omitted from the analysis.
The overall uncertainty in the line fluxes is obtained by taking into account the calibration error, the fitting error in the Gaussian, and the statistical error in the measurement itself.
\subsubsection{Scattered Light Evaluation}\label{S:stray4}
The EIS optics causes the instrument to scatter the
radiation coming from the solar disk into the detector, which can contaminate the observations even in the off-limb section of the slit. This contribution depends on the specific configuration of the instrument and on its pointing at the time of the observations and it cannot be removed a-priori. \citet{landi2007} devised a method to estimate the contribution of scattered light to coronal emission lines using concurrent observations of chromospheric lines or continuum emission. The presence of emission from chromospheric lines in off-limb observations is only due to scattered light, and its rate of decrease with height can be used to estimate its contribution to the total measured emission. In the case of the EIS spectrometer there is no continuum emission available. The only chromospheric line is from He II. \citet{Hahn2012} showed that the emission by this line in the off-limb section is actually real coronal emission, so this line cannot be used. EIS measured some transition region lines from O IV and O V which can potentially be used, but they are too weak. Instead, we evaluate the scattered light contribution based on EIS observations performed during a partial lunar eclipse. Using the flux ratio from the occulted and non-occulted portions of the disk, the EIS scattered light was found to be around 2\% of the disk emission (Ugarte Urra 2010, EIS Software Note No. 12). \par

\begin{figure}[h] 
\epsscale{1}
\plotone{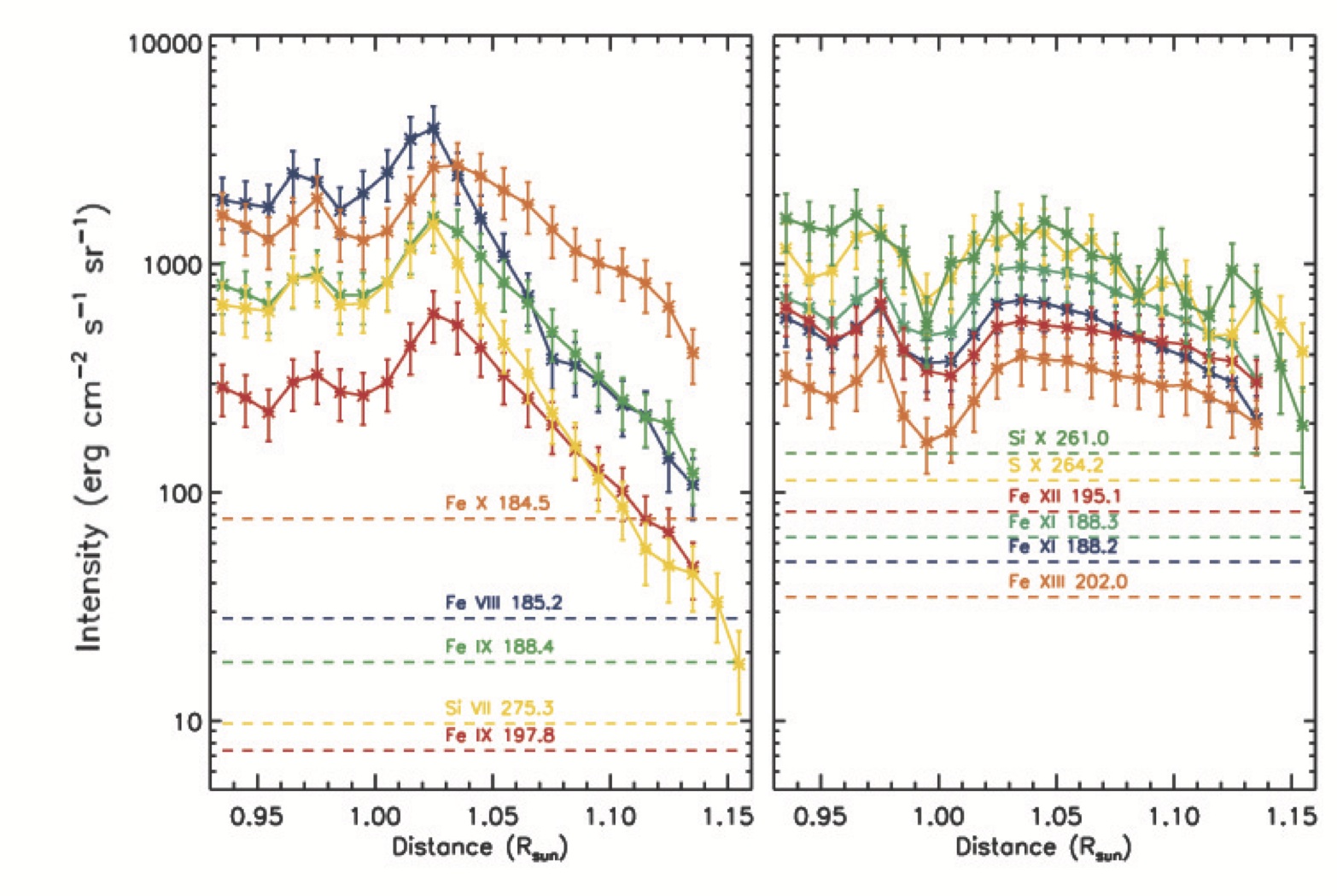} 
\caption{\small \sl Intensity vs. distance for the spectral lines in Table \ref{T:linesEIS}, over the EIS field of view between 0.93R$_\odot$ and the farthest end of the slit at 1.16R$_\odot$ (solid curves). The dashed lines show the estimated scattered light intensity for each line. The observed intensities and the scattered light level are color-coded in the same way. For clarity of presentation, the Si X intensity is multiplied by 10, S X by 12, and Fe XI 188.2 by 0.6. \label{F:stray4}}  
\end{figure}

We evaluate the scattered light flux for each of the lines in Table \ref{T:linesEIS} by averaging their emission in the portion of
the slit that covered the disk in the $0.61-0.97$R$_\odot$ range. The scattered light intensity is then taken to be 2\% of the average value. The line intensities over the EIS field of view from $0.93$R$_\odot$ to the end of the slit are shown in Figure \ref{F:stray4}. For clarity of presentation, the Si X intensity is multiplied by 10, S X by 12, and Fe XI 188.2 by 0.6. It can be seen the intensity drops sharply in the off-limb portion of the slit for the lines belonging to the lower ionization stages. This is consistent with having a small contribution from scattered light: in fact, the local coronal emission, which is proportional to $N_e^2$, decreases very rapidly with height from the limb, while scattered light usually decreases very slowly. The scattered light levels for each line are shown as dashed horizontal lines, and their values are reported in the third column of Table \ref{T:linesEIS}. These values should be taken as estimated upper limits, while the actual contribution is probably lower; in the present observations only part of the slit pointed into the solar disk, and therefore the telescope is less illuminated by the disk emission. To exclude any significant contamination by scattered light from this analysis, we conservatively use only observations where the estimated scattered light level is less than 20\% of the observed flux. The maximum heights at which this occurs for each of the lines, R$_{max}$, are reported in the last column of Table \ref{T:linesEIS}. \par  

\section{Results}\label{s:results}
\subsection{Solar Wind: Frozen-in Charge States}\label{s:results_uly}
\begin{figure}[p]
\epsscale{0.9}
\plotone{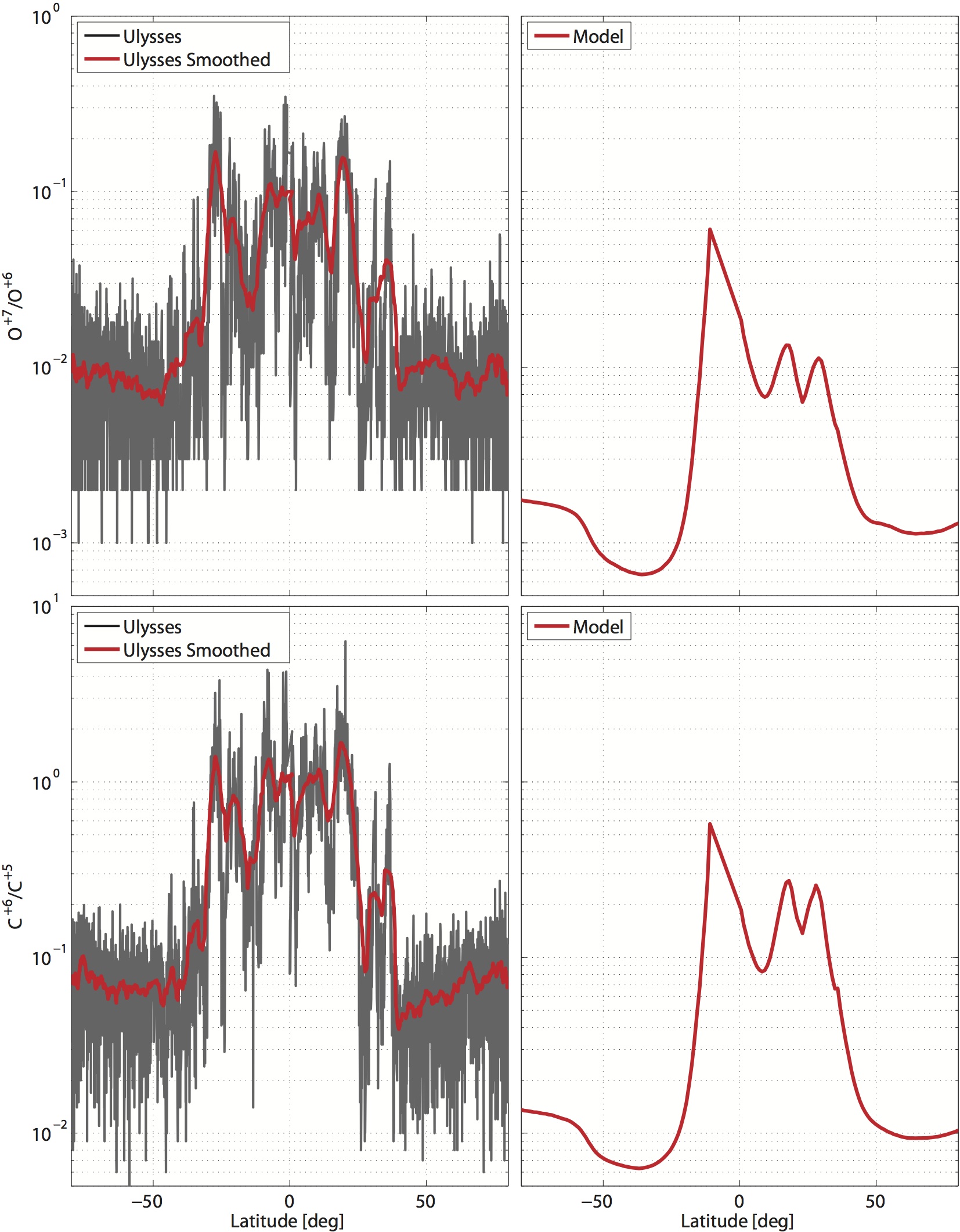}
\caption{\small \sl  Model-observation comparison of charge state ratios vs. heliographic latitude. The top and bottom panels show the comparison for $O^{+7}/O^{+6}$  and $C^{6+}/C^{5+}$, respectively. Left:  the gray curve shows Ulysses measurement taken at 3-hour intervals. The red curve shows the same data smoothed over a 6-day window. Right: ratios predicted by AWSoM-MIC for the field lines described in Section \ref{S:fieldlines}, plotted against the latitude reached by the field line at 1.8AU.\label{F:o7o6}}
\end{figure}

\begin{figure}[ht] 
\epsscale{0.9} 
\plotone{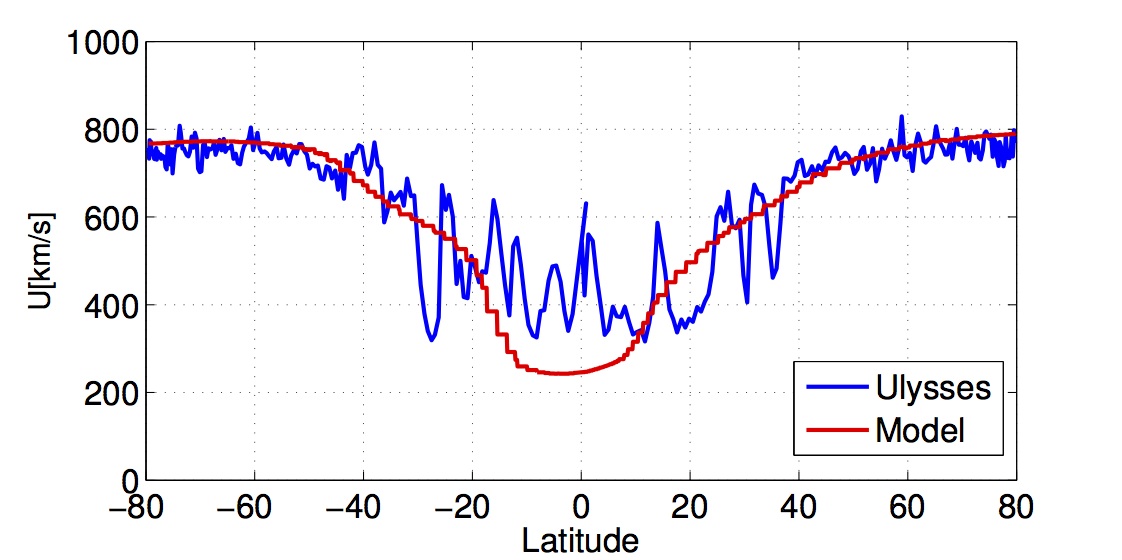} 
\caption{\small \sl Wind speed vs. heliographic latitude. The blue curve shows Ulysses measurements. The red curve shows the AWSoM result.\label{F:uly_model_ur}}
\end{figure}

The AWSoM-MIC frozen-in ratios from the field lines described in Section \ref{S:fieldlines} are compared to Ulysses observations in Figure \ref{F:o7o6}. The top and bottom panels show the comparison for $O^{7+}/O^{6+}$ and $C^{6+}/C^{5+}$, respectively, plotted against heliographic latitude. The left column shows the Ulysses observations, where the gray curve shows the original data at 3 hour resolution, and the red curve is a moving-average over a window of 6 days. The right column shows the corresponding AWSoM-MIC results for the case of a single temperature electron population. The first thing of note is that the predicted charge state ratios in the region around the equatorial plane are higher than those outside this region, in line with observations. This region corresponds to the location of the slow wind, as can be seen Figure \ref{F:uly_model_ur}), which shows the modeled (red curve) and measured (blue curve) speeds vs. latitude. The overall magnitude of the  modeled $O^{7+}/O^{6+}$ and $C^{6+}/C^{5+}$ ratios is about an order of magnitude lower than the observed values at all latitudes. However the qualitative behavior is markedly similar. 
The modeled charge states exhibit the well-known behavior of higher charge state ratios at low latitudes around the heliospheric current sheet, compared to lower (by about an order of magnitude) charge state ratios at high latitudes associated with polar CHs \citep{vonSteiger2000}.\par
Both ratios exhibit larger fluctuations when measured in the slow wind. This behavior cannot be addressed by our steady-state simulation, which cannot describe fluctuations anywhere. On larger time scales, the observations exhibit mid-scale variations on top of the overall variation between the fast and the slow wind. Similar behavior is seen in the model; however, as explained in Section \ref{S:fieldlines}, a simulation of a single Carrington Rotation can only be regarded as a ``snapshot" taken during Ulysses's polar scan, and the mid-scale variations seen in the model should not be directly compared to specific structures seen in the observations.\par 

These results demonstrate that fast and slow solar wind streams flowing along static open magnetic field lines can carry distinctly different frozen-in charge states. This result will be discussed in detail in Section \ref{s:ocb}.  The overall level of ionization we found in the simulation is too low at all latitudes. From Eq. (\ref{eq:mic}) we can see that insufficient ionization rates can be due to several factors: 1. the AWSoM electron density is too low, inhibiting the collisions necessary for ionization to the higher charge states ($C^{6+}$ and $O^{7+}$), or 2. predicted ionization rate coefficients are too small (which implies the thermal energy of the electrons is not predicted correctly), or 3. the ions flow speed below the freeze-in height is not predicted correctly, changing the time the different ions spend at each heights, and preventing sufficient ionization from occurring. We will explore these factors separately.\par 
\subsubsection{Modeled Electron Density and Temperature as a Cause of Under-predicted Charge States}\label{S:cause_density}
The coronal electron temperature and density predicted by the present simulation for CR2063 were validated in \citet{oran2013} using two sets of observations. First, they showed that the 3D thermal structure predicted for CR2063 leads to synthetic full-disk  images in the EUV and soft X-ray range (emitted by the lower corona) that are consistent with observations. Even though the discrepancy between the synthetic and observed full-disk images is larger at certain localized regions (especially around active regions), the large scale structure is well-reproduced. Second, the authors found that the modeled electron density and temperature at the center of the north polar CH were in good agreement with spectroscopic measurements extending from 1.05R$_\odot$ - 1.13R$_\odot$ above the limb.\par 
However, determining the electron density and temperature from remote observations is inherently complicated by line of sight effects, since the emission from different regions contribute to the measured intensity. \citet{Frazin2005, Frazin2009} and \citet{Vasquez2010} have developed a tomographic method to reconstruct the 3D thermal structure of the lower corona. The technique, dubbed \emph{differential emission measure tomography (DEMT)}, uses multi-wavelength EUV images of the lower corona taken from different points of view in order to reconstruct the electron density and temperature that are responsible for the emission. If a single observatory is used, the images are collected over an entire solar rotation, until a full coverage of the corona is achieved. For this reason DEMT can only recover steady structures; in regions where the magnetic topology or thermodynamic properties vary significantly during the rotation, the tomographic method fails to reconstruct a single set of thermal properties. These regions are excluded from the analysis. However, the global, large scale distribution can be reliably recovered.  In DEMT, the inner corona (1.02-1.20 R$_\odot$) is discretized on a regular spherical grid, with voxels having a radial size of 0.01 R$_\odot$ and angular size of $2^\circ$, both in the latitudinal and azimuthal directions. The tomographic 3D reconstruction of the EUV filter band emissivity in each band (Frazin et al. 2009) allows us to derive the \emph{local differential emission measure} (LDEM) in each voxel, which describes the distribution of temperatures of the plasma contained in that voxel. By taking moments of the LDEM, the final products of DEMT are 3D maps of the electron density, $N_e$, and the average electron temperature $<T_e>$ in each voxel of the tomographic grid. \par
\begin{figure}[ph] 
\epsscale{.9} 
\plotone{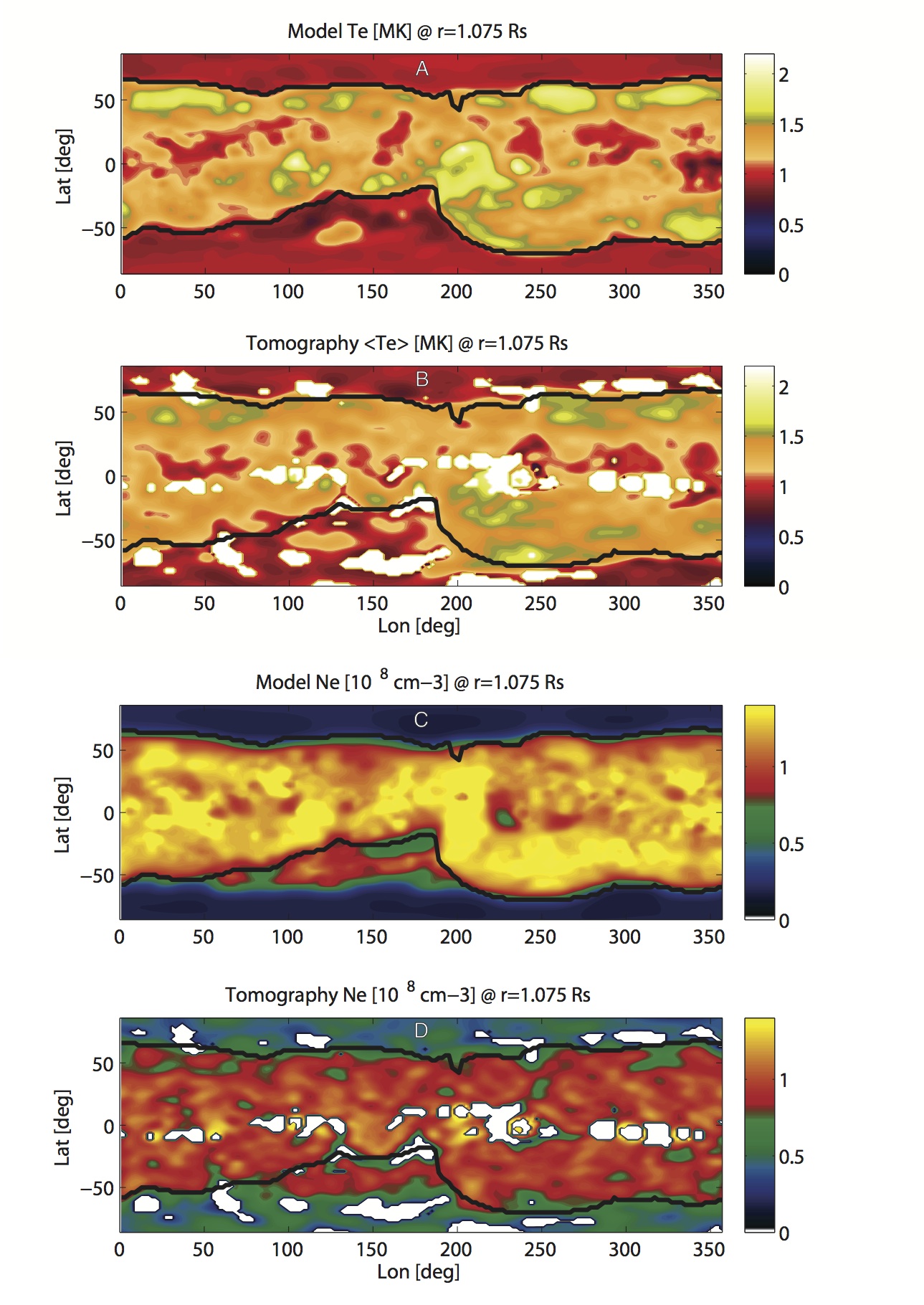} 
\caption{\small \sl  Model and DEMT maps for CR2063 extracted a height of 1.075R$_\odot$. Top two panels: AWSoM electron temperature $T_e$ and average electron temperature $<T_e>$ from DEMT. Bottom two panels: AWSoM electron density and DEMT electron density.  Black curves show the polar coronal hole boundaries extracted from the AWSoM solution. The white regions in the tomographic maps correspond to regions which could not be reconstructed by DEMT. \label{F:tom_te}}
\end{figure}

 
We performed a DEMT reconstruction for CR2063 using full disk images taken at three wavelengths by the Extreme Ultraviolet Imager (EUVI) on board the two STEREO spacecraft \citep{howard2008}. Figure \ref{F:tom_te} shows how the model compares to the reconstructed electron temperature and density. The data are plotted as a longitude-latitude map over a spherical surface extracted from the tomographic grid at $r=1.075$R$_\odot$. The top two panels show the comparison of modeled and tomographic electron temperature, while the bottom pair shows the same comparison for electron density. White regions in the tomographic maps correspond to regions where the tomography method fails, which occurs mostly around regions with high variability. The black curves show the boundary of the polar CHs based on the magnetic field from AWSoM. The mid-latitude regions, where the temperature and density are much higher, correspond to the closed field streamer belt.\par
The modeled CH boundaries follows the contours of the streamer belt in the tomography very closely, with small (up to 2-3 degrees) departures at certain regions. The open-closed boundary of the magnetic field is only plotted for polar CHs, but other closed field regions appear as islands of higher density and temperature outside the main streamer belt, while low-latitude CHs, having lower temperatures and densities, can be seen inside the main streamer belt. These regions have similar sizes and locations in both the model and the tomography. This comparison suggests that the magnetic field topology derived from the MHD solution at this height is realistic. Some discrepancies between the shapes of the CH boundary in the model and the tomographic density structure may be attributed to the fact that both the synoptic magnetogram used as a boundary condition to the model, and the tomographic reconstruction, were obtained from observations taken over the entire Carrington Rotation, and small scale and dynamic features will not necessarily be captured by either of these methods. \par 
While the modeled electron temperature is in very good agreement with the reconstructed values, the density comparison shows larger discrepancies, with the modeled density about 1.4 times larger than the reconstructed density in the closed field region, and about a factor 2 lower than the reconstructed density in CHs. \par

\begin{figure}[ht]
\epsscale{.50} 
\plotone{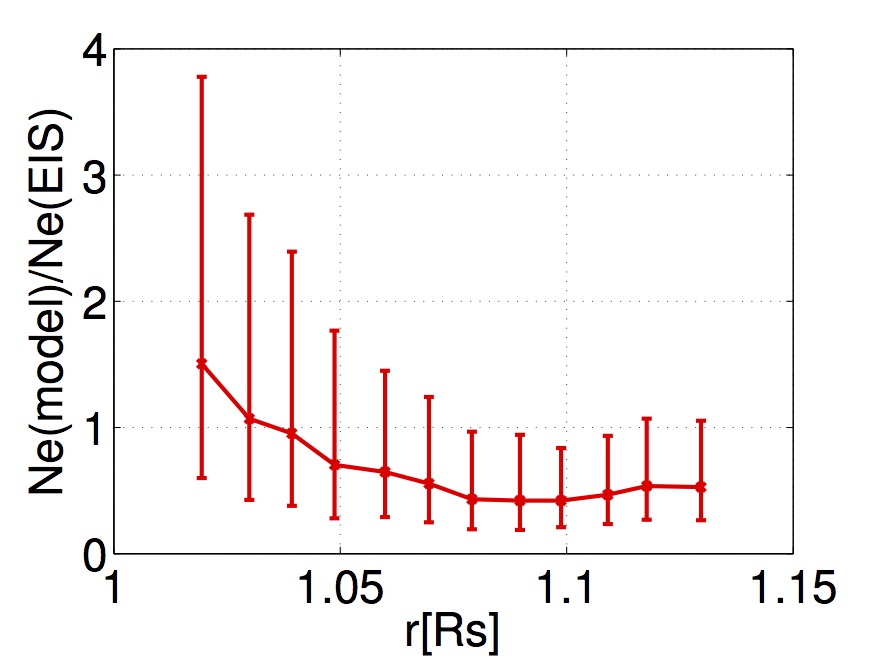} 
\caption{\small \sl The ratio of modeled to measured electron density vs. radial distance along the center of the north coronal hole. The electron density was measured using Fe VIII line intensities ratios measured by EIS.\label{F:eis_ratio}} 
\end{figure}

This under-prediction of the electron density in CH is also present at larger heights. Using the Fe-VIII line intensity ratios observed by EIS during CR2063, \citet{oran2013} measured the electron density along the center of the north CH, at heights between $1.02$R$_\odot$ - $1.13$R$_\odot$ above the limb, and compared them to model results (see Figure 13 therein). To make the comparison more quantitative, we calculate the ratio of modeled to measured density using the same data as in \citet{oran2013}. Figure \ref{F:eis_ratio} shows the ratio plotted against radial distance. The error bars are due to the uncertainty in the density measurements. Given these uncertainties, it is clear that the modeled values are within the uncertainties in the measurement at most heights. We note that the model/measured ratio is centered around 0.5 at heights $r>1.04$R$_\odot$, consistent with the model-tomography comparison.\par
The lower density predicted by AWSoM in the polar CHs would in general lead to lower collisions rate, and therefore to lower ionization. However, it is not immediately clear by how much an electron density that is a factor 2 too low would contribute to the under-prediction of the frozen-in values in Figure \ref{F:o7o6}, which are about an order of magnitude too low at all latitudes. To make a quantitative estimation, we repeated the charge state calculation for a few representative field lines, while multiplying the AWSoM electron density by a factor 2 at all points. We found that the resulting frozen-in values increase by about a factor 2. We conclude that the modeled electron density alone is not responsible for the difference between Ulysses and AWSoM-MIC charge state ratios.

\subsubsection{Impact of Supra-Thermal Electrons on the Ionization Rate Coefficients}\label{S:in_situ_super}
A second cause of under-predicted charge states is ionization rate coefficients that are too low. The rate coefficients depend on the thermal energy of the electrons. In solving Eq. \ref{eq:mic2}, we assumed the electron posses a Maxwellian distribution function, and calculated the rate coefficients based on the Maxwellian temperature. However, there could be additional thermal energy present, in the form of a supra-thermal tail of the distribution function. Even a small population of supra-thermal electrons can increase the ionization rate coefficients significantly. We therefore repeat the charge state calculations using ionization and recombination coefficients based on a main electron population obeying a Maxwellian at the modeled electron temperature, and an additional supra-thermal electron population, obeying a second Maxwellian at 3MK, which constitutes 2\% of the entire electron population. The values we used here to characterize the supra-thermal population were chosen for demonstration purposes only.  A more rigorous determination of these parameters requires exploring the parameter space through modeling and comparison to observations, and is beyond the scope of the present paper. We note these values are consistent with those used by previous authors, as discussed in Section \ref{s:supertheory}.\par
\begin{figure}[ph]
\epsscale{0.9} 
\plotone{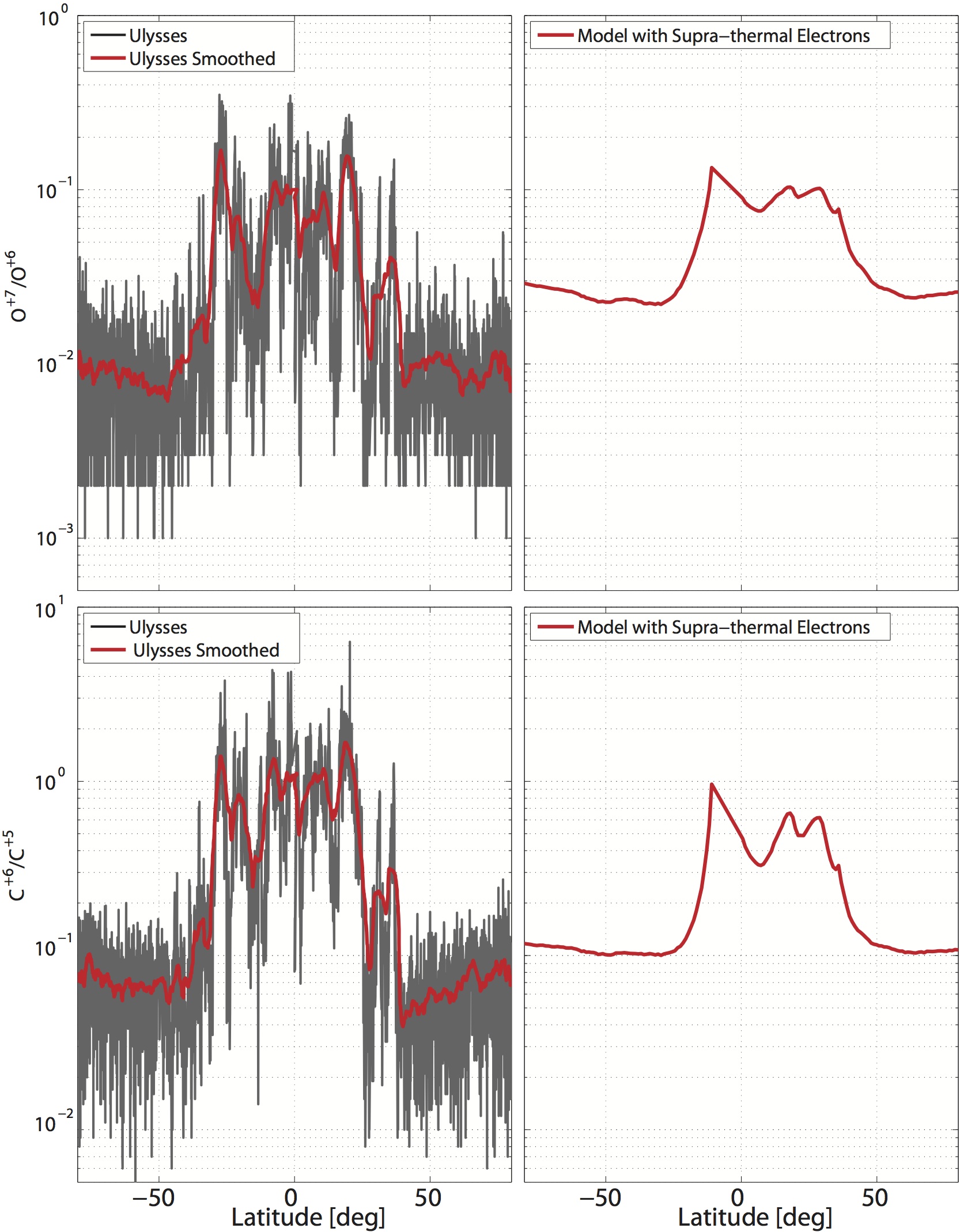} 
\caption{\small \sl  Model-observations comparison of charge state ratios vs. heliographic latitude, as in Figure \ref{F:o7o6}, but for the case where a supra-thermal electron population is added in the MIC simulation.\label{F:o7o6_super}}
\end{figure}

 
The results are shown in Figure \ref{F:o7o6_super}, with the same layout and color-coding as in Figure \ref{F:o7o6}. The agreement between the observed and predicted charge state ratios is significantly improved compared to the case without supra-thermal electrons. The modeled $C^{+6}/C^{+5}$ ratio is now in good agreement with the observations in both the slow and fast wind. This result is consistent with previous studies \citep[e.g., ][]{Esser2000, Laming2007} which showed that supra-thermal electrons can help solve the apparent discrepancy between observed and predicted charge state ratios in the solar wind. For the modeled $O^{+7}/O^{+6}$  ratio, the addition of supra-thermal  electrons allowed us to obtain a good agreement with observations in the slow wind, while in the fast wind this caused the ratio to become about a factor of 2-3 too high (compared to about an order of magnitude too low without the supra-thermal electrons). This suggests that further fine tuning of the supra-thermal populations size and energy is needed, before a truly accurate and acceptable agreement is obtained. This type of parameter search can be assisted by creating synthetic emission using the predicted ions fractions, to be compared with observations of the lower corona, as we present in Section \ref{s:results_eis}.\par 
 It is important to note that even though the supra-thermal electrons improved the agreement with the overall magnitude of the observed charge state ratios, they play no role in determining the large scale structure of these observables. In fact, the highest charge states occur at the same latitudes whether or not supra-thermal electrons are included, and they are increased by the same factor compared to the fast wind values (about one order of magnitude). Therefore, some other mechanism must be responsible for the higher charge states predicted in the slow wind, as will be discussed in detail in Section \ref{s:ocb}.\par 

\subsubsection{Ion Speeds as a Cause of Under-predicted Charge States}
A third cause for under-predicted frozen-in charge states may be due to an inaccurate prediction of the ion flow speeds. If the ion speed is so high that its travel time is shorter than the ionization time, the ionization to the higher charge states will be inhibited. There are two possible factors that can lead to ions speeds that are too high in the AWSoM-MIC simulations: either the wind speed itself is too high, or the assumption that all the ions move at the wind is wrong. We note here that the terminal wind speed in the model is in good agreement with Ulysses observations, especially in the fast wind (see Figure \ref{F:uly_model_ur}). However, it is still possible that the rate of acceleration at lower heights is not predicted correctly, affecting the evolution. This will be discussed further when we examine the charge state distributions in the lower corona in Section \ref{s:results_eis}.\par 
Alternatively, heavy ions can move at different speeds with respect to the background plasma, commonly referred to as differential flows. \citet{Burgi1986} showed that heavy elements, including C and O, should have flow speeds that are smaller than the proton speed at $r<20$R$_\odot$. \citet{Ko1997} showed that if the heavy ions move slower than the wind, higher ionization states are achieved, leading to a better agreement with in-situ observations.  It is also possible that ions of the same element flow at different speeds with respect to each other. \citet{Esser2001} showed that if ions with charge state m+1 flow faster than the ions with charge state m, then the recombination of the m+1 ions back to the m charge state can be significantly inhibited, resulting in higher ionization compared to a single-speed case. However,  the extent at which differential flows occur is not clearly known. One could hope to determine their extent empirically by changing the flow velocities of the different ion species until a good agreement with charge state observations is reached. However, Esser et al. (1998) found that the observed frozen-in charge state distributions could be reproduced by many different flow profiles, making it difficult to make a conclusive determination.  Furthermore, the effect of differential flows on the predicted charge states was found to be comparable to the effects of supra-thermal electrons \citep{Ko1998,Esser2001}. In fact, it is possible that both processes take place in the corona, and it is hard to determine their separate contributions. Here, again, the simultaneous comparison to in-situ and remote observations of the lower corona could assist in constraining parametric studies. 

\subsection{Lower Corona: Emission by Heavy Ions in a Polar Coronal Hole}\label{s:results_eis}
We calculated the synthetic LOS fluxes for all the lines in Table \ref{T:linesEIS} and compared them to their corresponding EIS observations. The magnitude of the synthetic emission from each point along the LOS is proportional to the relative abundance of the ion responsible for the emission, or the ion fraction, $N(X^{+m}) / N(X)$, as seen in Eq. (\ref{eq:GNT}). For each spectral line, we use ion fractions derived from:\\
\indent 1. charge state evolution in a single-temperature electron thermal core population.\\
\indent 2. ionization equilibrium in a single-temperature electron thermal core population.\\
\indent 3. charge state evolution assuming an additional supra-thermal electron population. \\
\indent 4. ionization equilibrium assuming an additional supra-thermal electron population. \\
Cases 1-2 and cases 3-4 will be based on different ionization and recombination rate coefficients (see Section \ref{s:supertheory}). Within each pair, the charge states are either allowed to evolve freely according to Eq. (\ref{eq:mic2}), or ionization equilibrium is imposed at each point along the trajectory (determined from the steady-state solution of Eq. (\ref{eq:mic})). This will allow us to gauge the extent of departures from equilibrium due to the flow speed. In what follows, we refer to the evolved charge states as MIC ion fractions. In cases 3-4, which include the supra-thermal electrons, we calculated the synthetic emission using modified level populations, as outlines in Section \ref{s:supertheory}.\par
\begin{figure}[hp]
\plotone{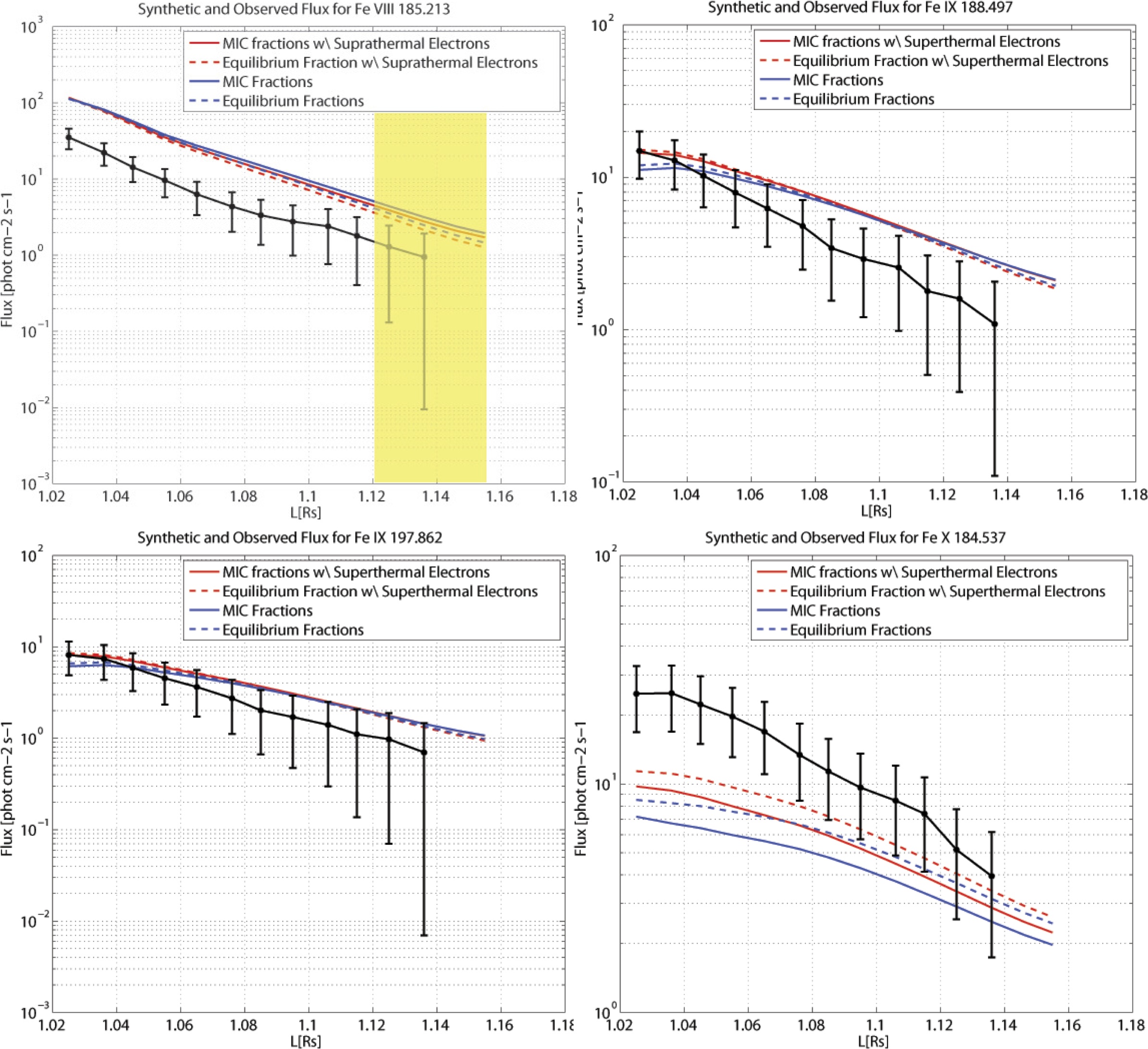} 
\caption{\small \sl  Observed and synthetic line of sight flux vs. radial distance for emission lines from Fe VIII to Fe X. The black curve shows EIS observations and their uncertainties. The two blue curves show the synthetic flux for a single-temperature electron population. The two red curves show the synthetic emission including supra-thermal electrons. In each pair, the solid curve was obtained using the MIC ion fractions in the contribution function, while the dashed curves were obtained using ion fractions determined from ionization equilibrium. The shaded area represents heights at which the scattered light may contribute more than 20\% to the observed flux. \label{F:fe_08_10}}
\end{figure}

\begin{figure}[hp] 
\plotone{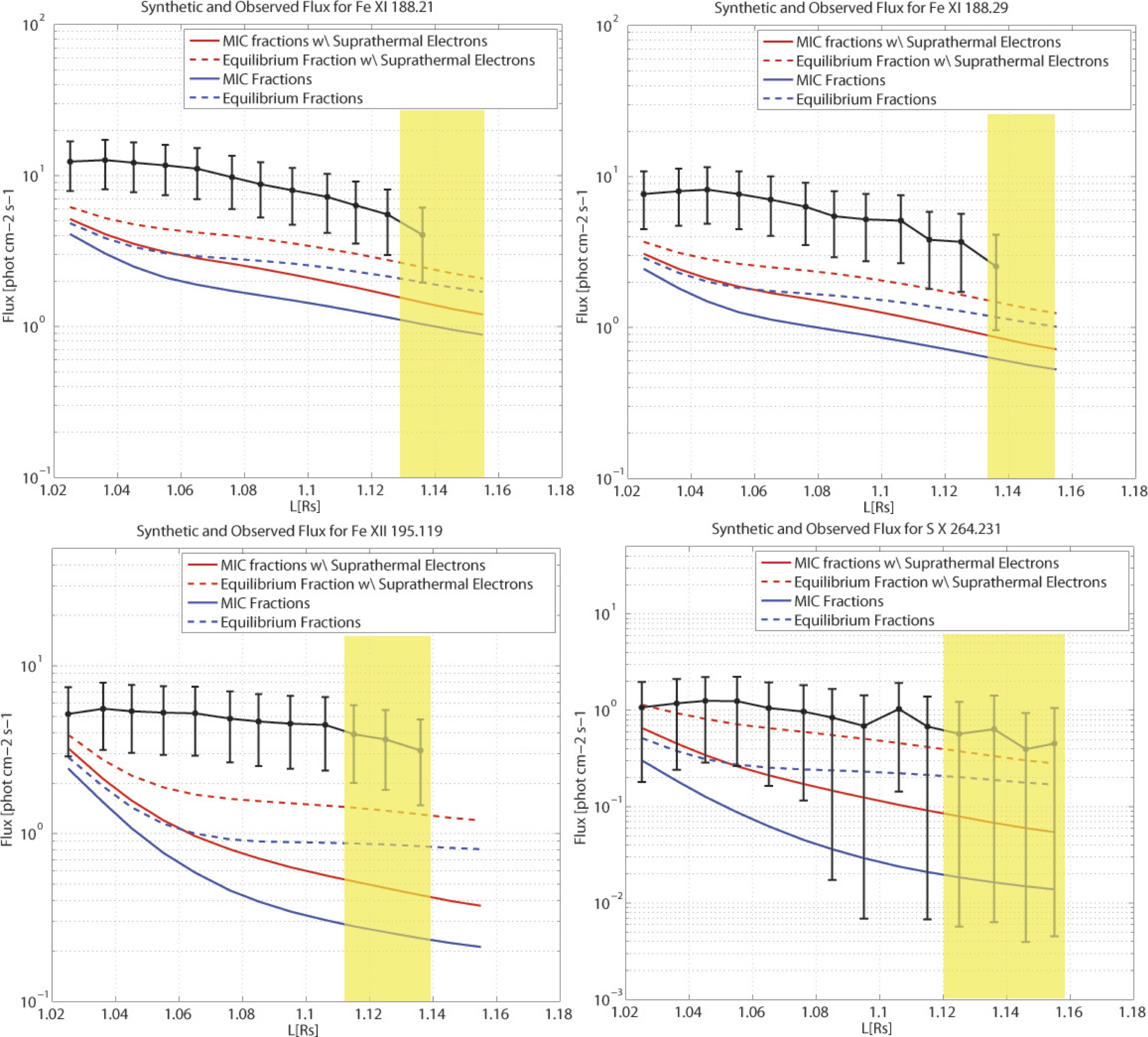} 
\caption{\small \sl  Observed and synthetic line of sight flux vs. radial distance for emission lines from Fe XI, Fe XII and S X. The color coding is similar to Figure \ref{F:fe_08_10}.\label{F:fe_11_12_S_10}}
\end{figure}

 
Figures \ref{F:fe_08_10} - \ref{F:si_7_10} show the comparison of the synthetic and EIS fluxes as a function of height for all the lines. In each figure, the black curve shows the EIS observations and their uncertainties. The two blue curves show the synthetic flux for a single-temperature electron population, while two red curves are for the supra-thermal case. Within each pair, the solid curve is based on MIC ion fraction, while the dashed curve is based on ionization equilibrium fractions. The height ranges shaded in yellow represent the distances at which scattered light contamination may be higher than 20\% of the observed flux, taken from Table \ref{T:linesEIS}.\par 
\subsubsection{Under- and Over- Predicted Charge States}\label{S:over pop}
There are 7 lines covering different charge states of Fe, from 8 to 12. As can be seen, the synthetic emission is over-predicted for charge states 8 and 9, while it is under predicted for charge states 10-12, for all four types of predicted ion fractions. The best agreement is achieved for spectral lines belonging to Fe IX 197.862 \AA, where the synthetic emission is within the uncertainty of the measured  flux at most heights. \par 
The fact that the synthetic fluxes are either over- or under-predicted for ions of the same element, removes the possibility that the disagreement is due to uncertainties in elemental abundances, as these should shift all the predicted fluxes in the same direction. Another source for the discrepancy could be contamination from  hotter streamer material that might cross the line of sight, which will preferentially contribute to the observed emission from the higher charge states. This contribution is hard to quantify from line of sight observations alone; however, the magnetic field configuration obtained by the model shows that no closed field lines cross the line of sight. The physical interpretation of these discrepancies  is that Fe is not ionized rapidly enough in the model, leading to an over-population (and emission) of low charge states and an insufficient population of high charge states. \citet{Landi2014} found similar behavior when analyzing synthetic emission from several models, including the AWSoM model, for an ideal dipole magnetic field case.\par
Since Fe only freezes-in around 4R$_\odot$, the model may still achieve the correct ionization status at altitudes higher than the EIS field of view, and specifically the correct frozen-in charge states.  To examine this, we compared the predicted frozen-in  value of $<Q>_{Fe}$ to the Ulysses observations made above the north polar CH, which is the other end of the wind trajectory for most of the plasma observed here by EIS. The results are shown in Figure \ref{F:qfe_north}. The gray curve shows the value of $<Q>_{Fe}$ measured by Ulysses/SWICS at 3-hour resolution vs. latitude. The blue curve shows a moving average over a 6-day window, while the red curve shows the modeled frozen-in values (for the case including supra-thermal electrons). It can be seen that the modeled $<Q>_{Fe}$ is very close to the observed values, and it differs by less that one charge state from the smoothed values. Recalling that the charge state composition has an uncertainty between 10-25\%, it is clear that the discrepancy between the model and the observations at Ulysses's orbit is small compared to that found in the lower corona; there, the emission from the highest charge state in our data set, Fe XII, is almost an order of magnitude lower than the observations, even with the inclusion of supra-thermal electrons. Thus we can conclude that the under-predicted ionization of Fe in the lower corona eventually recovers at larger heights, at least partially, giving rise to frozen-in values that closer observations. \par
\begin{figure}[hp]  
\plotone{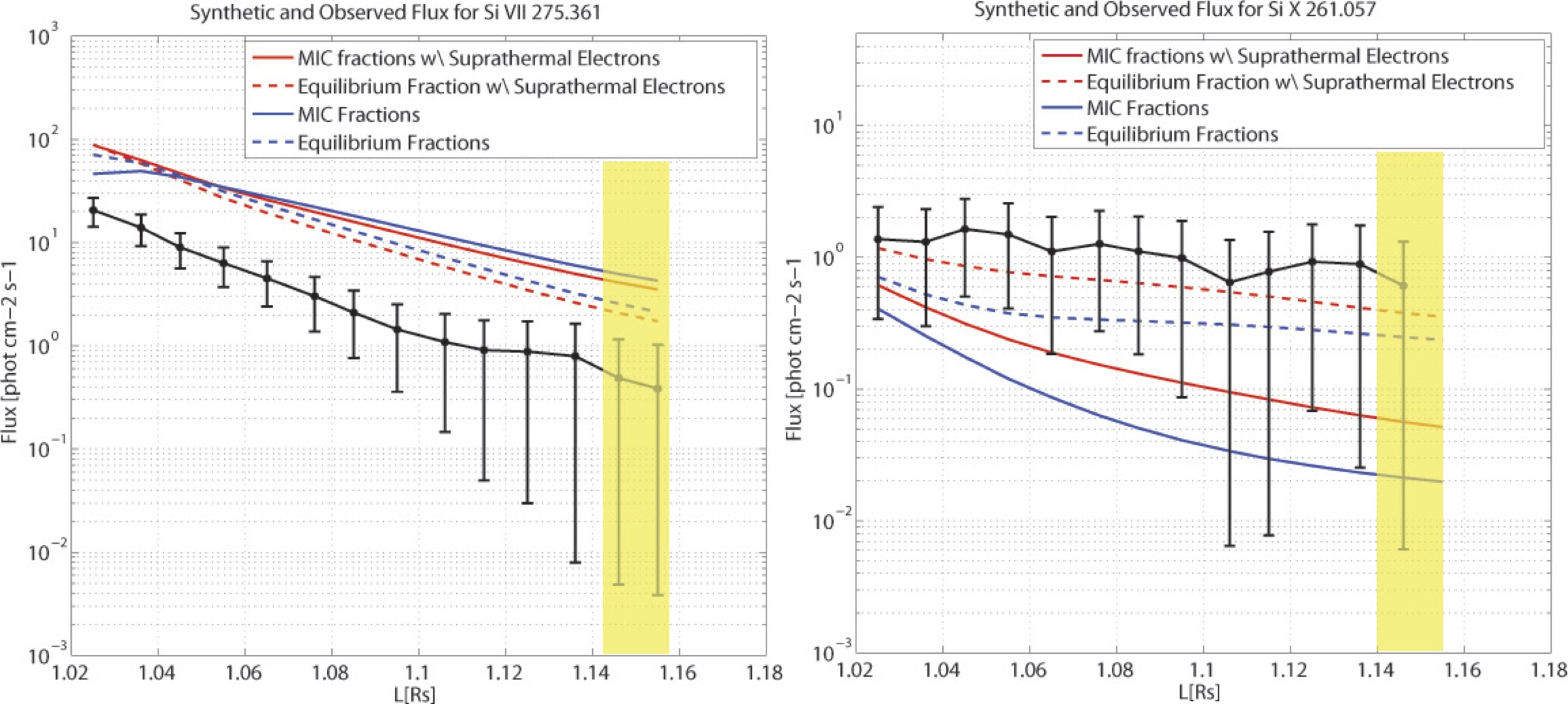} 
\caption{\small \sl  Observed and synthetic line of sight flux vs. radial distance for emission lines from Si VII and Si X. The color coding is similar to Figure  \ref{F:fe_08_10}.\label{F:si_7_10}}
\end{figure}

\begin{figure}[hp]
\epsscale{0.6}
\plotone{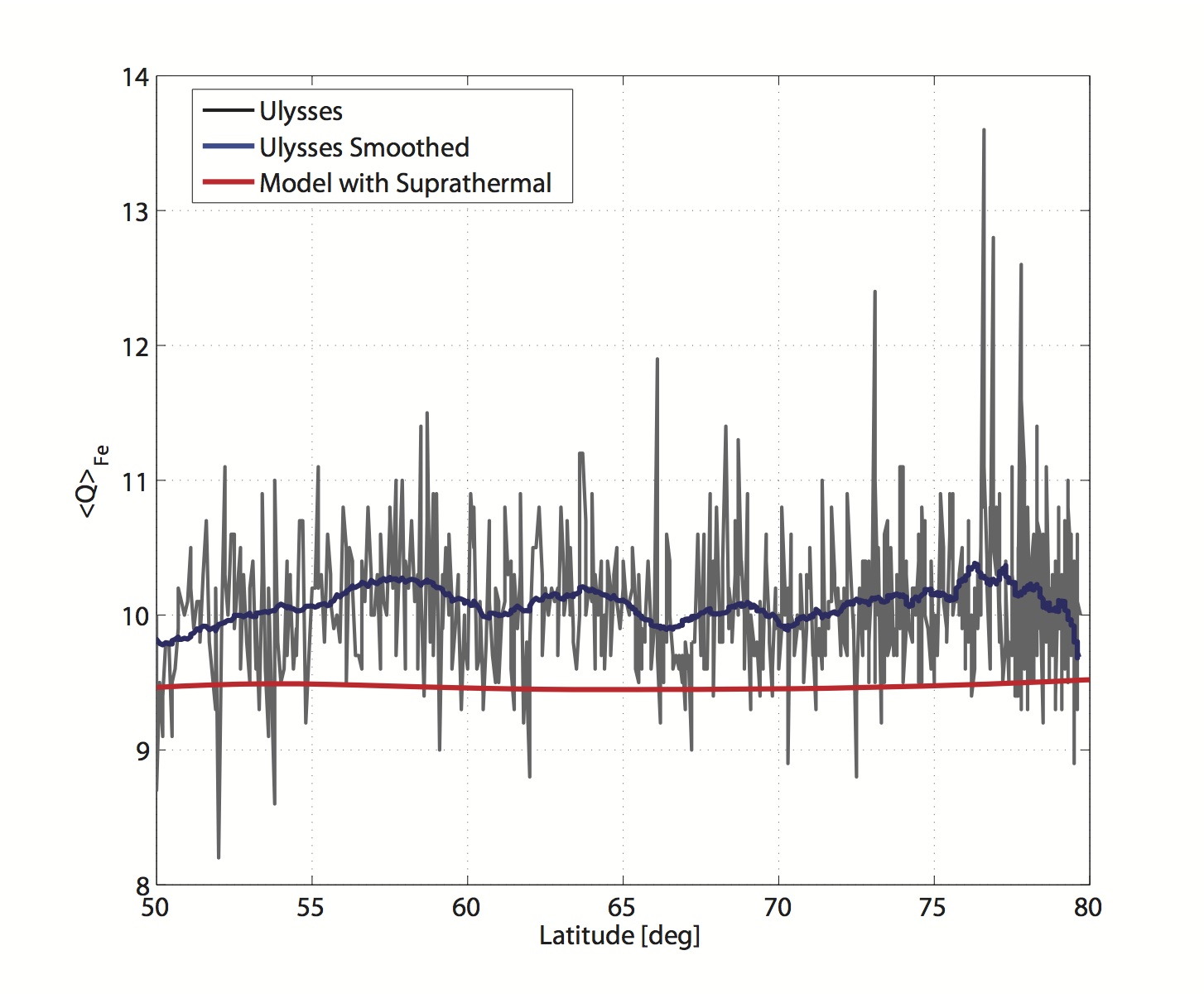}[ht]
\caption{\small \sl  The frozen-in average charge state of Fe plotted vs. heliographic latitude above the north coronal hole. The gray curve shows Ulysses measurement taken at 3-hour intervals. The blue curve shows the same data smoothed over a 6-day window. The red curve shows the average charge state predicted by AWSoM-MIC (for the case including supra-thermal electrons).\label{F:qfe_north}}
\end{figure}

 
The same effect can be seen in the two lines belonging to Si (Figure \ref{F:si_7_10}), where the Si VII line flux is over-predicted and that from Si X is under predicted. Unfortunately there are no publicly available data of Si charge states from Ulysses at the time of this publication. Finally, the agreement between the predicted and observed flux for the S X line is very good. However, since only a single line is used here, it cannot reveal further information about the charge state evolution.\par

\subsubsection{Spectral Signatures of Supra-thermal Electrons}\label{S:super_results}
In many of the spectral lines the supra-thermal electrons give rise to a noticeable difference in the predicted fluxes, making this type of model predictions a potential diagnostics for the properties of the supra-thermal electrons themselves.
In these lines, the inclusion of supra-thermal electrons improved the agreement between predicted and observed values. The fluxes emitted by the low ionization states, which are over-predicted, are smaller in the supra-thermal case,  while the reverse occurs for the under-predicted fluxes from the higher charge states. This can be explained by the fact that the supra-thermal electrons increase the ionization rate coefficients; in this case a larger portion of the element is ionized to a higher charge state, leaving less ions in the lower charge states. The resulting emission from the low and high charge states decreases or increases, respectively, becoming closer to the observed values for all charge states.\par 
This result, taken in conjunction with the comparison of modeled and observed frozen-in charge states discussed in Section \ref{s:results_uly}, demonstrates that supra-thermal electrons below the freeze-in height lead to a better agreement with observations at both ends of the wind trajectory.  Furthermore, by calculating the emission assuming a non-Maxwelian electron distribution function, we showed that supra-thermal electrons may have a spectral signature. This serves as a proof of concept that the presence of supra-thermal tails below the freeze-in height may reconcile the discrepancies between the coronal electron temperature derived from spectral observations (which usually assume a Maxwellian electron population) and the temperature required to produce the frozen-in charge states. A better agreement with the observations can be achieved by empirically adjusting the parameters of the supra-thermal electron populations, i.e. their relative portion of the entire population, and their energy. Such a procedure can help pin-down the properties of supra-thermal electrons by attempting to reproduce the emission from as many lines and from as many instruments as possible. However, the spatial  distribution of supra-thermal electrons may not be uniform below the freeze-in height, as pointed out by \citet{Laming2007}. This introduces additional degrees of freedom in any parametric study aiming to determine the properties of supra-thermal electrons.\par 
\subsubsection{Departure from Equilibrium and Wind Acceleration}
The synthetic emission calculated using equilibrium ion fractions agrees better with the observations compared to the MIC ion fractions, both with and without supra-thermal electrons. In other words, the model over-estimates the departures from equilibrium. This may be explained by ion speeds that are too large, not allowing them sufficient time to achieve a charge state distribution that is closer to the equilibrium for the local conditions. An over-predicted wind speed is also consistent with the over-population of the low charge states of Fe and Si, which occur for both ionization equilibrium and for fully-evolved charge state distributions, as discussed in Section \ref{S:over pop}.\par
As in the case of the in-situ charge states, these discrepancies might be resolved if the ions are allowed to have differential flow speeds, in effect changing the ionization rates. Another possibility is that the predicted wind speed is not realistic. We saw that the wind speed at 1-2AU agrees well with the observations, especially above the CH (see Figure \ref{F:uly_model_ur}); however, it may still be too large below the freeze-in height. If this is the case, then it implies that the wind acceleration process assumed in the model might need to be further refined. In AWSoM the wind is accelerated by gradients in the Alfv{\'e}n wave pressure, the thus the wave reflection coefficient will have a large impact on the wind acceleration rate. In the AWSoM simulation used in this work, taken from \citet{oran2013}, the authors assumed a spatially uniform reflection coefficient. In reality, the reflection coefficient depends on the gradients in the Alfv{\'e}n speed, and thus it will vary with location. Future work will explore these effect using a self-consistent description of the reflection coefficient, as the one presented in  \citet{vanderholst2014}.

\section{Discussion: The Highly Ionized Steady Slow Wind}\label{s:ocb}
The main result of Section \ref{s:results_uly} is that the observed large-scale variation of the charge state ratios $O^{7+}/O^{6+}$ and $C^{6+}/C^{5+}$ with latitude can be produced by a model where both fast and slow wind come from coronal holes and flow along static open magnetic field lines. This is an important result, since the slow wind charge states often serve as observational support to dynamic release models, in which the source region of the slow wind are coronal loops, and the acceleration mechanism is driven by intermittent reconnection events. It is therefore worthwhile to understand how the variation in charge states between the steady fast and slow wind is obtained by the model, which is the subject of the present Section.\par
Before we attempt to answer this question, it is important to put this work in context. A steady state model cannot describe any transient phenomena, and thus cannot address the high and sudden fluctuations of the charge states observed in the slow wind; these are probably caused by dynamic release due to reconnection between open and closed field lines \citep{Fisk1998, Fisk2003, FiskZhao2009,Wang2000,Antiochos2007,Antiochos2011,Antiochos2012}, as discussed in the Introduction. The AWSoM model also does not include a mechanism for heavy element fractionation, and therefore cannot address the FIP-bias found in the slow wind. Thus, our results cannot be used to contradict the dynamic release models. Rather,  they offer a complementary picture to slow wind formation, as they demonstrate that a sub-class of slow wind can exist that does not come from coronal loops, and which carries high ionization levels that are already skewed toward the typical values observed in the slow wind, albeit without the fluctuations. If this is indeed the case, this type of slow wind will be relatively steady, will carry high charge states, but most likely will not exhibit a FIP-bias, since biased abundances are generally formed in closed-field structures \citep[e.g. ][]{Feldman2003}. The relation between this complementary picture and dynamic release models will be discuss in more detail in Section \ref{S:implications}.  
\subsection{The Source Region of the Steady Slow Wind}\label{S:region}
\begin{figure}[p] 
\epsscale{0.59}
\plotone{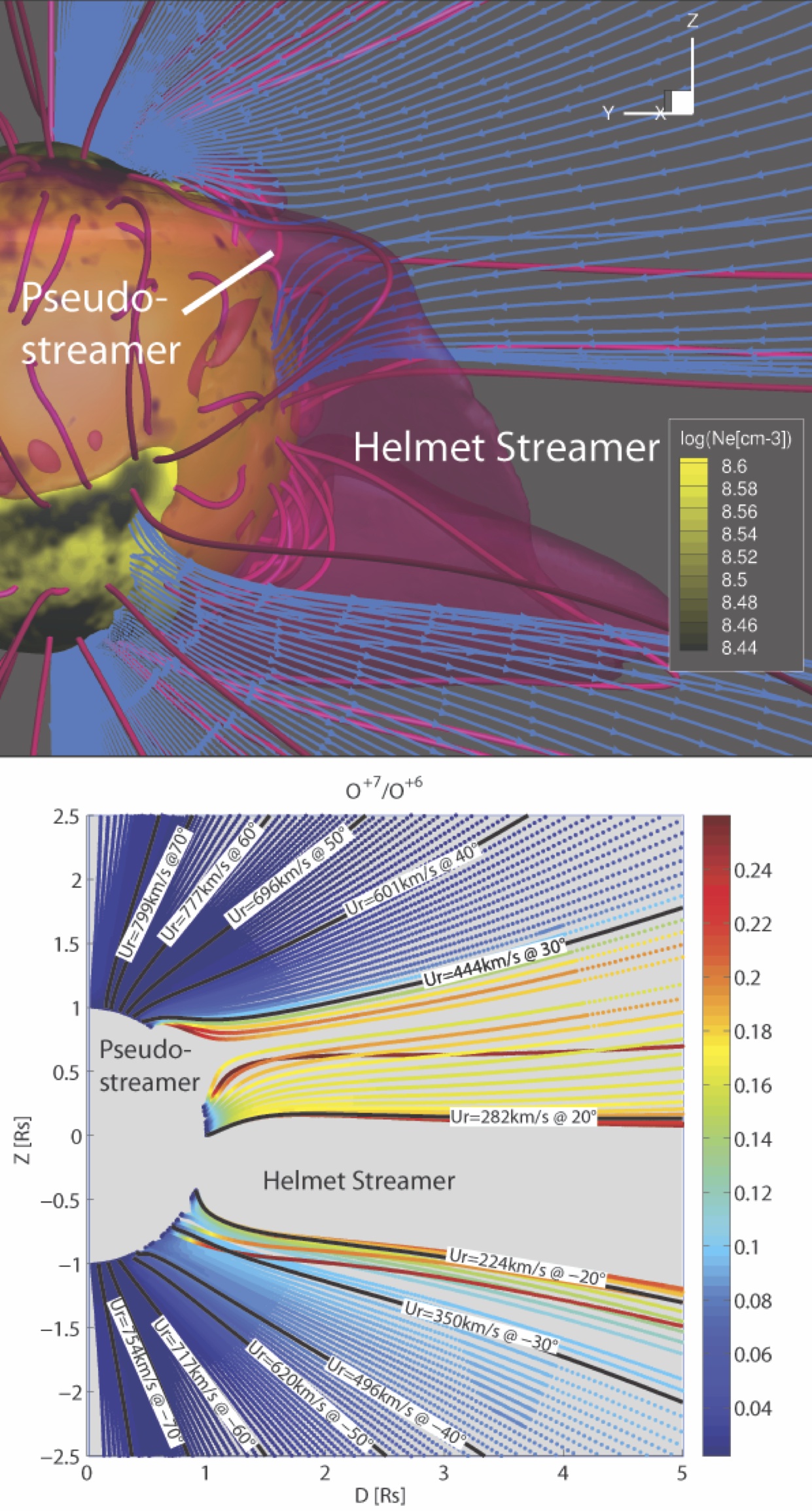} 
\caption{\small \sl  Top: AWSoM solution for CR2063. The solar surface is colored by the electron density.  Blue curves show open magnetic field lines for which the charge states evolution is analyzed in Section \ref{S:region}. Purple curves show selected open and closed magnetic field lines. The purple surface is an iso-surface of electron temperature at 1.6MK, showing the general shape of the helmet streamer. Bottom: Predicted $O^{7+}/O^{6+}$ ratio along the blue field lines shown in the top panel, presented in one plane. The field lines are colored by the local charge state ratio. Black field lines are those reaching 1.8AU at 10 degree spacing in latitude. The labels show the wind speed and latitude at 1.8AU of the respective line.\label{F:o7o6_analyze}}
\end{figure}

The latitudinal variation of the frozen-in charge state ratios seen in the AWSoM-MIC results suggest that the open field lines carrying the the fast and slow wind undergo different evolution below the freeze-in height. In order to characterize these differences, and locate the source regions of the different wind types, we examine the evolution of the charge states and wind properties close to the Sun. We choose a new set of open field lines with foot points locations ranging from the poles toward the streamer belt, in both hemispheres. These are shown as the blue curves in the top panel of Figure \ref{F:o7o6_analyze}. The solar surface is colored by the electron density, while thick purple lines show additional open and closed field lines, representing the overall structure of the corona. The purple transparent surface is an iso-surface of electron temperature at 1.6MK, which shows the general shape of the streamer belt. To make the analysis simple, we selected field lines that are rooted close together in longitude, so that the conditions encountered by adjacent field lines will vary smoothly. The mean longitude of the foot points in the northern and southern hemispheres are different, due to the shape of the streamer belt separating the two groups.
For the northern hemisphere group, the open field lines belong to three different structures, from north to south: a polar CH, a pseudo-streamer, and a low latitude CH just below it. For the south hemisphere, the selected field lines come mostly from inside the polar CH, but their foot points extend into lower latitude than the north hemisphere group, where they straddle the boundary of the helmet streamer from the left.\par
The bottom panel of Figure \ref{F:o7o6_analyze} shows the same blue field lines shown in the top panel, flattened onto one plane for clarity, where the vertical and horizontal axes represents the distance from the equator and the distance from the polar axis, respectively, of each point along each field line. The field lines are colored by the magnitude of $O^{7+}/O^{6+}$ predicted by an AWSoM-MIC simulation with supra-thermal electrons. The labeled black field lines demonstrate how the magnetic field in the corona maps to the heliosphere: the ends of these field lines intersect a spherical surface at 1.8AU at $10^\circ$ spacings. The labels show the wind radial speed and the heliographic latitude at that distance. The regions covered by the helmet streamer and the pseudo-streamer are also labeled. Note that the range of attitudes without open field lines only reflects the structure close to the Sun; out in the heliosphere, these latitudes will be filled by field lines rooted in other longitudes on the solar surface.\par 
The distribution of $O^{7+}/O^{6+}$ in Figure \ref{F:o7o6_analyze} shows that the highest charge state ratios ($\sim$0.2) originate from the pseudo-streamer and the low-latitude CH just below it, and are carried by a slow wind. Charge state ratios of $\sim$0.1 originate from the edges of the polar CHs, and are also carried by slow wind flows (up to $450$km s$^{-1}$). These field lines reach latitudes of up to $\pm 40^\circ$ at 1.8AU. In contrast, the fast wind ($>600$km s$^{-1}$) comes from deeper inside the polar holes and carries charge state ratios between 0.02-0.08, smoothly increasing from the center of the hole toward lower latitudes. These values are consistent with those used by \citet{Zurbuchen2001} to distinguish between fast and slow wind streams in in-situ observations taken during solar minimum. Using Ulysses and ACE data, they found that the slow wind exhibited ratios at and above 0.1, while values of $O^{7+}/O^{6+}<0.1$ were associated with fast wind streams coming from polar CHs. \citet{Zurbuchen2002} found that the polar fast streams can carry $O^{7+}/O^{6+}$ lower than 0.02, which is similar to the lower limit of  the frozen-in $O^{7+}/O^{6+}$ ratio we found in simulating this specific set of field lines.\par 

The connection we made in Figure \ref{F:o7o6_analyze} between the wind at 1.8AU and the corona reveals three source regions of highly ionized slow wind streams: pseudo-streamers, low latitude CHs, and the boundaries of polar CHs. The latter was suggested to be the source region of the slow wind by several authors, who related the low speeds to the larger expansion of the open flux tubes rooted this region \citep{Suess1979,Kovalenko1981,Withbroe1988,Wang1990,cranmer2005}. \citet{cranmer2007} calculated the charge state evolution of O ions in an axially symmetric solar model driven by turbulent waves. Their model prescribed an idealized magnetic field topology of expanding flux tubes, where the expansion factor increased from the center of the CH toward the streamer leg. They found that the resulting frozen-in charge state ratio $O^{7+}/O^{6+}$ increases with decreasing wind speed, which is in qualitative agreement with the observations. However, inside the fast wind, their predicted charge state ratio showed a sharp increase when moving from wind speeds of $\sim$650km s$^{-1}$ toward $\sim$750km s$^{-1}$ (i.e. toward the center of the CH). This increase, amounting to around an order of magnitude in size, is not in agreement with the Ulysses observations, and may be due to the assumed magnetic field topology. Here, we have directly simulated the charge state evolution at all latitudes using a realistic magnetic configuration, and verified that the observed charge state ratios can be reproduced with values that are in agreement with observations, at least in their large scale behavior.\par
In summary, in the AWSoM-MIC simulations, the coronal hole boundaries form the low latitude slow wind, which carries charge states of about 0.1 for the case of $O^{7+}/O^{6+}$, while other open field regions such as the pseudo-streamer supply an even higher charge state ratio (around 0.2). Thus our simulations show that the steady-state model can not only produce higher charge states in the slow wind, but it can also account for some of their variations within the slow wind, which can be linked to the magnetic topology of the corona. This is a distinct capability of a global model that is constrained by the observed magnetic field.\par
 \subsection{How and Why are the High Charge States Formed?}\label{S:how}
\begin{figure}[p] 
\epsscale{0.49}
\plotone{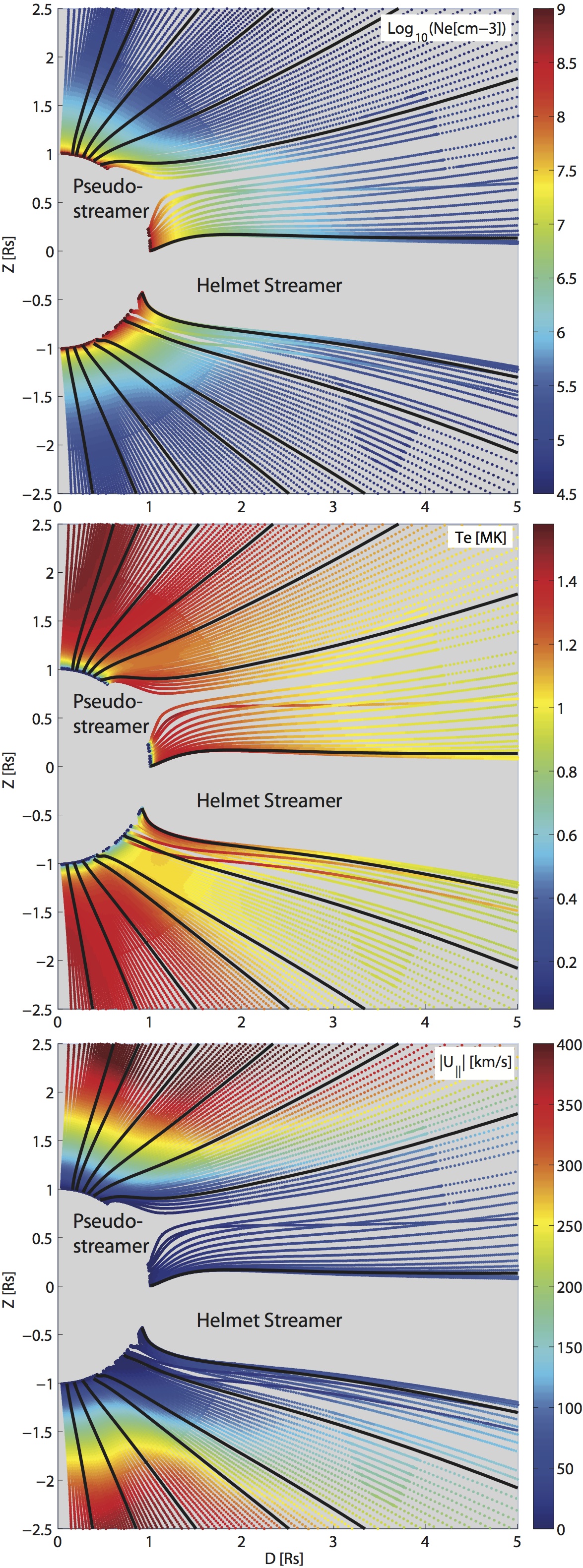} 
\caption{\small \sl  AWSoM solution for CR2063, along the blue field lines in the left panel of Figure \ref{F:o7o6_analyze}. From top to bottom: electron density, electron temperature and speed parallel to the field line. Black curves are the same as in the right panel of Figure \ref{F:o7o6_analyze}.\label{F:wind_analyze}}
\end{figure}

The ionization status of a given element at a given location along a field line depends on the wind condition along its path up to that point. As is clear from Eq. (\ref{eq:mic}), the properties that control the evolution are the electron density and temperature, and the wind speed. These quantities are plotted in Figure \ref{F:wind_analyze}, along the same field lines as in Figure \ref{F:o7o6_analyze}. The black field lines are identical to those plotted in Figure \ref{F:o7o6_analyze}, but their labels were removed for clarity. The top panel shows the electron density, the middle panel shown the electron temperature, while the bottom panel shows the wind speed.

In the previous section, we identified the polar coronal hole boundary (CHB) regions as the source region of a large part of the slow and highly ionized wind. The field lines belonging to the CHBs in Figure  \ref{F:wind_analyze} exhibit higher electron densities at their base, and a slower fall-off of density with radial distance, compared to lines coming from deeper inside the CHs (top panel). Examining the electron temperature (middle panel), we can see that the largest temperatures near the foot points occur very close to the streamer leg. The higher densities in the CHB (as well as in the pseudo-streamer and the low-latitude CHs) also lead to lower wind speeds (bottom panel) due to conservation of mass flux. Thus the CHBs are characterized by higher electron density, higher electron temperature, and slower wind speed compared to deeper in the coronal hole at any given height. The higher density and temperature would lead to higher ionization rates, which are proportional the electron density, and increase with increasing incident electron energy. Furthermore, due to the lower speeds, the CHB wind will spend more time in the collisional environment close to the Sun, allowing for more ionization to occur. All these factors combine to produce as overall higher ionization, and higher frozen-in charge states.\par
 It is interesting to note that the electron temperature above the poles, which are the source region of the fast wind, can be almost as high as to that reached along CHB field lines. Despite this fact, the fast wind does not get ionized to similar levels as the slow wind. This is because the density falls off faster in the fast wind, inhibiting collisions with electrons and causing the charge states to freeze-in before they reach the higher temperature regions along their trajectory. This points to an important limitation of methods that infer coronal temperatures from in-situ charge state observations: if the density is low enough in the lower corona, the frozen-in charge states will will not carry information about higher temperatures that may be reached above the freeze-in height.\par
 
 \subsection{The Steady Wind from CHBs as a Subset of the Non-Steady Slow Wind}\label{S:implications}
 The picture presented here of a the formation of a steady and highly ionized slow wind complements dynamic release models as follows. The Ulysses observations show that the mean level of charge state ratios is higher in the slow wind than in the fast wind (see, for example, the smoothed curve in  Figure \ref{F:o7o6_super}). Furthermore, charge state ratios as low as those found in the fast wind are rarely present in the slow wind observations covered in this data set. This pronounced increase in charge states is consistent with a scenario where the observed non-steady slow wind is in fact a mixture of material from closed field regions and material from the open field lines from the polar CHBs and low latitude CHs, which already carry charge state ratios that are higher than those observed in the fast wind. Thus it is possible that the slow wind simulated by the steady-state model can be a constituent of the variable non-steady slow wind. In this case, this subset of slow wind will be steady and will carry intermediate to high charge states. Since it does not originate from closed magnetic structures, we can expect it to have similar elemental abundances as that of coronal holes.\par
  This sub-set of the slow wind has been possibly identified in Ulysses/SWICS measurements of the solar wind by \citet{Stakhiv2014}. The future Solar Orbiter mission may allow us to further examine whether this wind can be detected in observations. This mission, due to launch on January 2016, will approach the Sun at distances as close as 0.28AU. The Heavy Ion Sensor (HIS), which is part of the Solar Wind Analyzer on board Solar Orbiter (Solar Orbiter Definition Study Report, 2011), will be able to measure the ionic charge states and abundances of key elements, offering a new window into the state of the solar wind before it is modified by its propagation through the complex structure of the heliosphere.\par 
  
 \subsection{Enhanced Electron Density and Temperature at the Source Region}
The formation of the highly ionized steady slow wind in the AWSoM-MIC simulations is explained by the fact that the electron density and temperature are higher and the wind speed is lower at the slow wind source regions compared to those found in the source region of the fast wind (see Section \ref{S:how}). For the picture to be valid, these properties of the source region of the slow wind have to be confirmed observationally, and, if possible, explained theoretically. Further, if the electron density and temperature enhancement are indeed responsible for the formation of the highly ionized steady slow wind, then they should be present globally, and not only in the set of field lines we analyzed in Sections \ref{S:region} and \ref{S:how}.
Since most of the slow wind comes from the polar CHBs, we will focus on these regions and defer the analysis of the more complex low-latitude CHs and pseudo-streamers to a separate study. 
\subsubsection{Observational Evidence using EUV Tomographic Reconstruction}\label{S:tomo}
\begin{figure}[ht]
\plotone{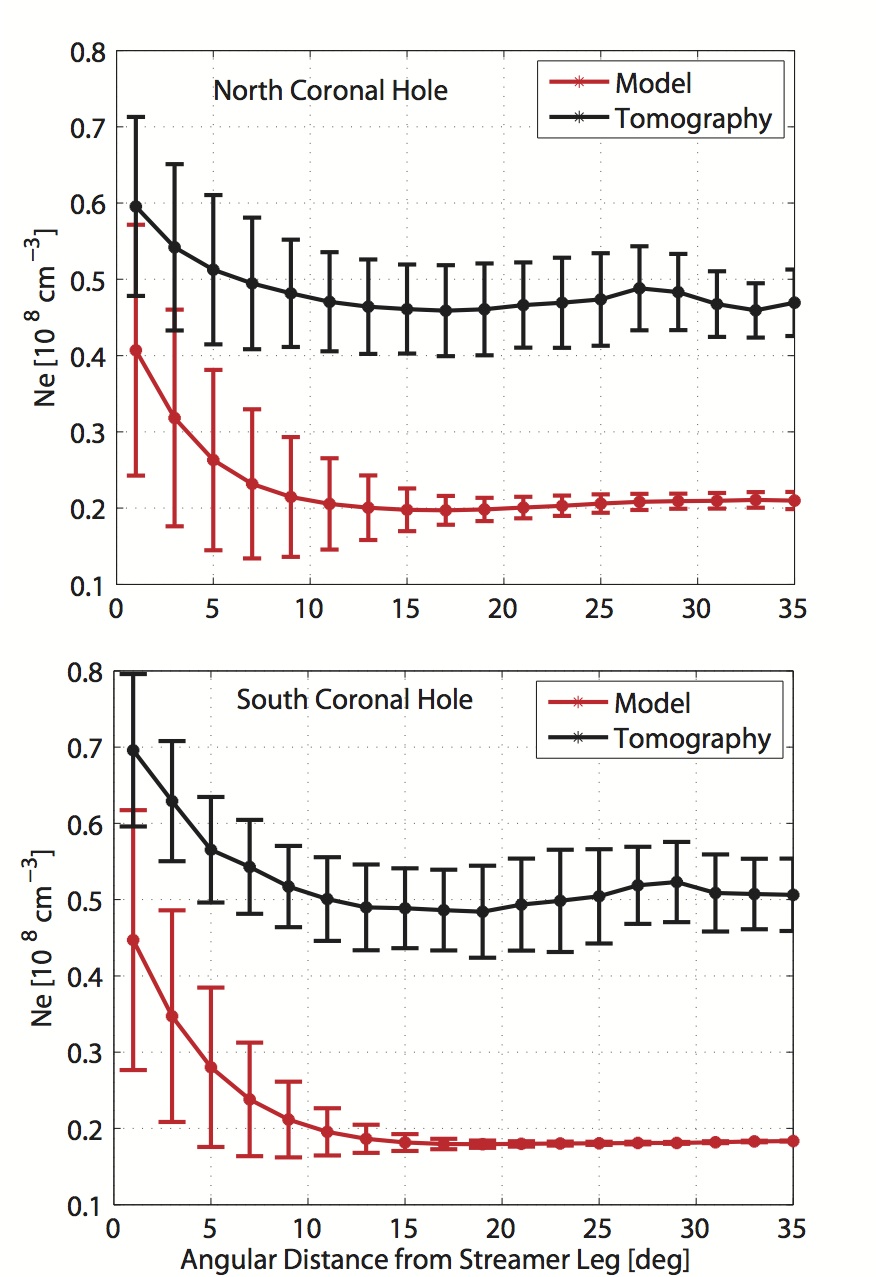} 
\caption{\small \sl Electron density vs. angular distance in the north (top) and south (bottom) coronal holes for CR2063, extracted from the model and tomography density maps at $r=1.075$R$_\odot$. Angular distance is measured from the streamer leg ($0^o$) toward the pole ($30^o$). The density is averaged over all longitudes. The black and red curves shows data extracted from tomography and the model, respectively. Error bars show the standard deviation from the averaged values taken from all longitudes. \label{F:ne_dist}} 
\end{figure}

We use the tomographic reconstruction of CR2063 presented in Section \ref{S:cause_density} to determine whether the CHB region exhibits the higher densities and temperatures predicted by AWSoM. It is hard to discern these properties just by inspecting the tomographic maps in Figure \ref{F:tom_te}. For a clear quantitative examination, we calculate the average variation of density with latitude over the entire polar CHs. For each longitude, we extract from the models and tomographic maps the electron density as a function of angular distance (in latitude) from the edge of the streamer belt towards the pole, where we define the edge of the streamer as the first open field line from the model, which appears as the black curves in the maps. For each angular distance, the densities from all longitudes are averaged together. A box in the longitude range of [50, 260] and latitude  [-90,30] was excluded from the analysis, since this region exhibits a large extension of the CH into lower latitudes, embedded with several islands of closed field regions. The results are shown in Figure \ref{F:ne_dist} for the north and south CHs. The black curve in each plot shows the density profile extracted from the tomography, while the red curve shows that extracted from the modeled density map. The error bars represent the standard deviation from the average over longitude. The modeled density is lower than the reconstructed density, by a factor of 2-3, which is expected since this discrepancy exists in the maps. However, two important features emerge in both the model and the tomography averages: \par
1. the density is highest at the edge of the CH, and smoothly decreases until it reaches an almost constant value by 10-15 degrees away from the outer edge.\par
 2. the rate of decrease vs. angular distance is similar in both the model and the tomography.\par
  In a DEMT analysis of the latitudinal dependence of the electron density during solar minimum, V\'asquez et al. (2010) found it to increase from the CH boundary towards the poles (see their Figure 6). In the present tomographic reconstruction we applied a blind-deconvolution of the point spread function (PSF) of the EUVI images, using the algorithm developed by \citet{Shearer2012}. The results shown in Figures 25 and 26 therein strongly suggest that density variation inside the open field region found by V\'asquez et al. (2010) were due to scattered light contamination. The use of the \citet{Shearer2012} algorithm effectively removes this contribution, and makes our conclusion that the density varies with latitude more reliable. \par
\begin{figure}[ht] 
\plotone{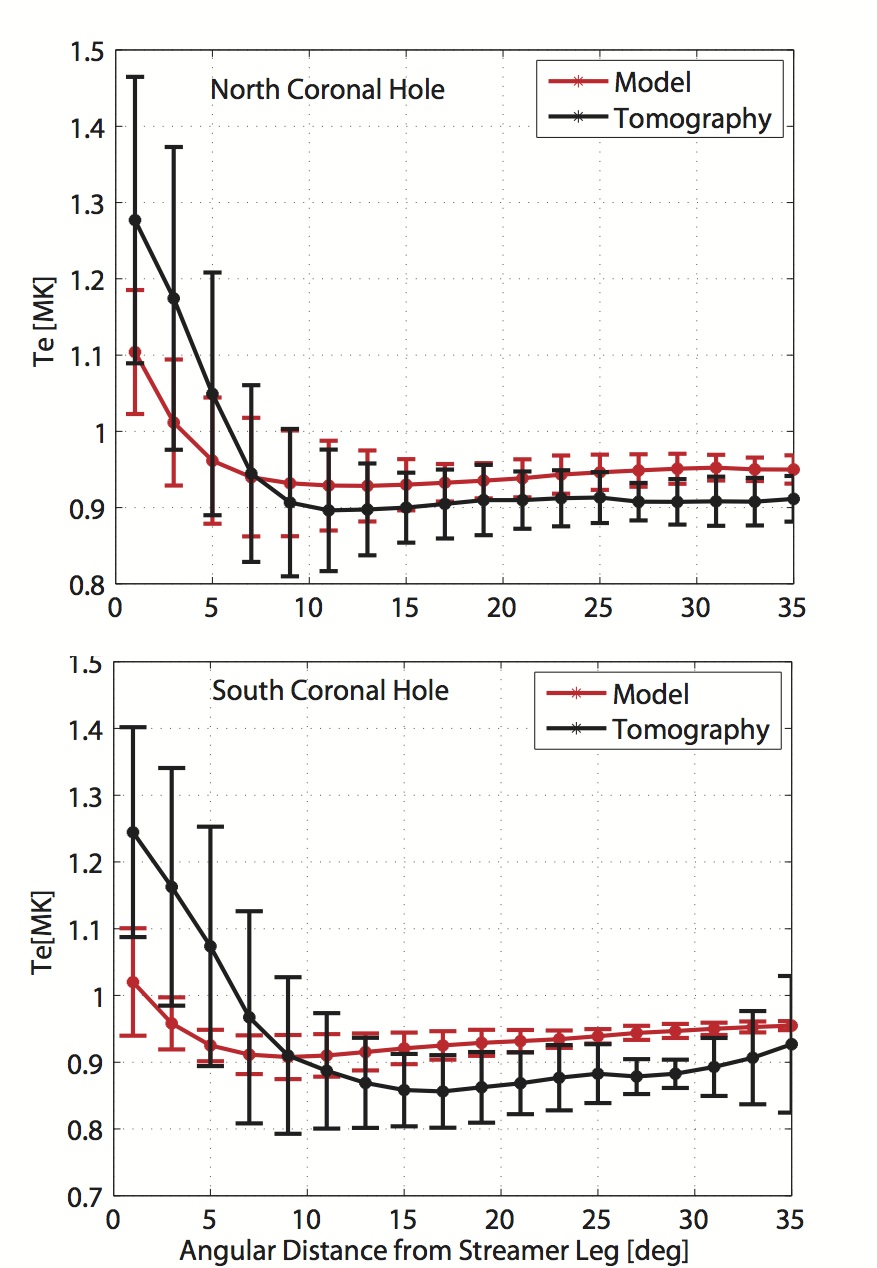} 
\caption{\small \sl Electron temperature vs. angular distance in the north (top) and south (bottom) coronal holes for CR2063, extracted from the model and tomography density maps at $r=1.075$R$_\odot$. Angular distance is measured from the streamer leg ($0^o$) toward the pole ($30^o$). The temperature is averaged over all longitudes. The black and red curves shows data extracted from tomography and the model, respectively. Error bars show the standard deviation from the averaged values taken from all longitudes. \label{F:te_dist}} 
\end{figure}

 
We next perform the same statistical analysis for the modeled and reconstructed electron temperature in the polar CHs. The variation of electron temperature as a function of angular distance from the CH edge is shown in Figure \ref{F:te_dist}, for the north (top panel) and south (bottom panel) CHs. The agreement between the model and the reconstructed values is good (as can be clearly seen in the tomographic maps themselves). Both the tomographic reconstruction and the model show that the electron temperature increases towards the edge of the hole. The model under-predicts the temperature in the CHB region, and the agreement improves as we move toward the poles.\par
In the previous section, we showed that an electron density and temperature enhancement in the CHB region in the lower corona is responsible for the increased charge states in the wind coming from this region. The analysis of the tomographic data confirms that such an enhancement is present on the Sun, and that this behavior is characteristic of the entire CHB region at all longitudes. It also shows that even though the model under-predict the absolute values in the CHB, it does correctly predict the variation with latitude of these quantities inside the CHB region.\par

\subsubsection{Theoretical Considerations}
The formation of enhanced electron density and temperature in the CHB region should be studied rigorously in order to obtain a consistent theoretical picture. This should involve more sophisticated simulations and observations than we used in this work. We here only offer possible conceptual explanations that should be further verified. The simplest explanation is related to the expansion of flux tubes. Those rooted in the CHB region will in general have a larger expansion factor compared to those rooted in the center of the CH. This can lead to two processes that can enhance the electron density. First, the larger expansion will lead to a slower wind coming from the CHB (as can be seen in the bottom panel of Figure \ref{F:wind_analyze}), which will in turn lead to higher densities. It is not clear, however, by how much it will affect the density at the very low height where the tomography maps were extracted ($r=1.075$R$_\odot$), as the wind speeds at these heights are very low.\par 
Second, flux tubes with larger expansion are magnetically connected to a larger volume of the hot corona. This may enable field-aligned electron heat conduction to transport larger amounts of thermal energy back to the chromospheric foot point. As a result, the energy per unit area reaching the chromosphere in the CHB region will be higher compared to deeper inside the CH. This may result in higher rates of chromospheric evaporation \citep[c.f. ][]{klimchuk2006}, whereby heated chromospheric plasma advects upward, supplying the coronal portion of the flux tube with denser material. This upward extension can be sustained in steady-state due to radiative cooling, which increases with the density and works to balance the every from heat conduction. In steady state, this will result in a nonuniform transition region, one that reached different heights for different flux tubes.\par
A variable transition region height can also occur due to variations in the Alfv{\'e}n wave Poynting flux, as demonstrated in \citet{suzuki2013}. These authors showed that changing the Poynting flux will also result in different fall-off of density with distance. This mechanism is balanced by radiative cooling in a somewhat similar way as in chromospheric evaporation. Finally, we note that all the effects above may contribute to the observed enhancement, and a more explicit study should be made to determine their relative contributions.\par
In the AWSoM model, since the density at the inner boundary is fixed, and thus it cannot respond to the heat conducted from the corona or to excessive wave heating. In order to determine how the model equations respond to these, a full time-dependent simulation with dynamic boundary condition is required. However, the inclusion of electron heat conduction and radiative cooling allows the model to mimic the phenomena described above. In a steady-state, the heating rate, which is the sum of the local wave heating rate and the heat transport from the corona, is balanced by the radiative cooling rate. As the latter is proportional to the square of the electron density, the steady-state solution will adjust the radial profiles of the electron density accordingly.\par

\section{Conclusions}\label{s:discussion_ch4}
The work presented here has combined, for the first time, results from a global 3D model of the solar atmosphere with a heavy ion evolution model, in order to simulate the large scale latitudinal structure of charge states in the corona and solar wind. Charge states have long been a key observational constraint for theories aiming to explain the processes responsible for the formation and acceleration of the fast and slow solar wind. Any such theory should also explain the observed variations in elemental abundances between the fast and slow solar wind, namely the appearance of the FIP bias in the slow wind abundances. The AWSoM-MIC simulation presented here cannot address the FIP bias, as the AWSoM model does not describe the separate evolution of the different species, and does not incorporate any fractionation mechanism. In addition, the steady-state simulation presented here cannot capture the observed variability in the slow wind properties. However, although this work cannot solve all the open question regarding the origin of the slow wind, understanding the large scale structure of charge states in the fast and slow solar wind provides an important piece of the puzzle. The capability to predict charge states from a global model using a realistic magnetic configuration is a major step forward in developing tools to test our understanding of solar wind formation and acceleration, and to ultimately predict space weather.\par 
The main result of this work is that we were able to produce higher levels of the frozen-in charge state ratios  $O^{7+}/O^{6+}$ and $C^{6+}/C^{5+}$ in the slow wind, compared to those in the fast wind without invoking release of material from the closed field region. We have shown that open flux tubes carrying higher charge state ratios are characterized by lower wind speeds and larger electron densities in the lower corona, where the electron temperature reaches its maximum. These field lines are rooted in a pseudo-streamer, a low-latitude CH, and in the boundary region between CHs and the streamer belt. The latter class of field lines are mapped to latitudes between $\pm 40$ in the heliosphere. This means that the boundary region in the model has a higher density compared to deeper inside the CH. The electron density and temperature enhancement was shown to be a global feature of CHs in the Carrington Rotation under question, both in the global model results, and in a tomographic reconstruction of the lower corona.\par 
The theoretical picture presented here of a steady slow wind coming from CHBs does not contradicts dynamic release models. Rather, they can be unified. The CHB lines in our steady-state simulation already carry charge state ratios that are consistent with the average level observed in the non-steady slow wind; however, the charge state ratio in the slow wind fluctuates rapidly and can reach values that were not captured by the simulation. Thus these larger charge state ratios can be due to reconnection of CHB lines with closed field lines at the edges of the streamer belt (a scenario similar to the S-web model presented in \citet{Antiochos2011, Antiochos2012}).  A possible prediction from the work presented here is that the CHB is the source region of a slow, steady, and highly ionized slow wind, but one that exhibits elemental abundances similar to those of CH and the fast wind, that is, without a FIP-bias. In an accompanying paper \citep[][under review]{Stakhiv2014}, this hypothesis was explored observationally by analyzing large amounts of in-situ data. \citet{Stakhiv2014} have shown that these is indeed a subset of solar wind flows with high charge states but no FIP bias.\par 
 The charge state distributions for Fe, Si and S below the freeze-in height were used to calculate synthetic emission that was compared to EIS observations in the lower corona, up to $1.115$R$_\odot$ above the limb of a polar CH. Comparing the results for 10 spectral lines suggests that the overall plasma ionization at this height range is too low; emission from low charge state ions was over-predicted while emission from higher charge states of the same ion was under-predicted. This suggest that the AWSoM wind profiles, and most probably the wind speed below the freeze-in height, need to be improved in order to reach a better agreement. The electron density is also under-predicted in CHs, and this also could cause the wind's ionization state to be lower relative to equilibrium.\par
We have explored the possible role that supra-thermal electrons can play in charge state evolution. Such an electron population has been hypothesized to be present in the corona, but no direct observational evidence of their existence has been found. We have shown that supra-thermal electrons at $\sim$3MK making  up 2\% of the entire electron population can greatly improve the agreement between the predicted and observed charge state levels in the solar wind, consistent with previous work \citep[e.g. ]{Ko1997, Esser2000}. \par
The addition of supra-thermal electrons also improved the agreement between the observed and synthetic fluxes of all of the 10 emission lines considered here. To the best of our knowledge, this is the first time a possible observational signature of the presence of supra-thermal electrons was found in remote spectral observations. This serves as a proof of concept for constraining our estimates of the energy and population size of supra-thermal electrons. Future work should include a parametric study, guided by observations at both ends of the wind trajectory, in order to pin down their properties.\par 
The AWSoM/MIC predictions can be improved by using a more sophisticated description of the solar atmosphere. For example, the wind speed below the freeze-in height can be improved by including a physics-based description of wave reflections \citep{vanderholst2014}. In addition, the effect of differential speeds of the heavy ions can be included by extending the two-temperature MHD description to a multi-fluid MHD description.\par

\acknowledgments
\begin{center}
\bf{Acknowledgments}\normalfont\\
This work was supported by the NSF grant AGS 1322543.
The work of E. Landi is supported by NASA grants NNX10AQ58G, NNX11AC20G, and 
NNX13AG22G. The simulations performed in this work were made possible thanks to the NASA Advanced Supercomputing Division, which granted us access to the Pleiades Supercomputing cluster.
Analysis of radiative processes was made possible through the use of the CHIANTI atomic database. CHIANTI is a collaborative project involving the following Universities: Cambridge (UK), George Mason and Michigan (USA).\par 
\end{center}



\clearpage 

\begin{thebibliography}{78}
\providecommand{\natexlab}[1]{#1}
\expandafter\ifx\csname urlstyle\endcsname\relax
  \providecommand{\doi}[1]{doi:\discretionary{}{}{}#1}\else
  \providecommand{\doi}{doi:\discretionary{}{}{}\begingroup
  \urlstyle{rm}\Url}\fi

\bibitem[Abbo et al.\ (2010)]{abbo2010}
Abbo, L., Antonucci, E., Miki{\'c}, Z., Linker, J. A., Riley, P., and
  Lionello, R. 2010, Advances in Space Research, 46, 1400--1408,
  \doi{10.1016/j.asr.2010.08.008}

\bibitem[Alazraki \& Couturier\ (1971)]{alazraki1971}
Alazraki, G., and P.~ Couturier 1971, \aap, {13}, 380

\bibitem[Antiochos et al.\ (2007)]{Antiochos2007}
Antiochos, S.~K., DeVore, C. R.,  Karpen, J. T.,  and Miki{\'c}, Z. 2007, \apj,
  671, 936--946, \doi{10.1086/522489}

\bibitem[Antiochos et~al.\ (2011)]{Antiochos2011}
Antiochos, S.~K., Miki{\'c}, Z.,  Titov, V. S., Lionello, R., and Linker, J. A. 2011, \apj, 731, 112, \doi{10.1088/0004-637X/731/2/112}.

\bibitem[Antiochos et~al.\ (2012)]{Antiochos2012}
Antiochos, S.~K., Linker, j. A., Lionello, R., et al. 2012, \ssr, 172, 169--185,
  \doi{10.1007/s11214-011-9795-7}.

\bibitem[Antonucci et~al.\ (2012)]{antonucci2011}
Antonucci, E., Abbo, L., and Telloni, D. 2012, \ssr, 172, 5--22, \doi{10.1007/s11214-010-9739-7}.

\bibitem[Asplund et~al.\ (2009)]{Asplund2009}
Asplund, M., Grevesse, N., Sauval, A. J., and Scott, P. 2009, \araa, 47, 481--522,
  \doi{10.1146/annurev.astro.46.060407.145222}.

\bibitem[Banerjee et~al.\ (2011)]{banerjee2011}
Banerjee, D., Gupta, G. R., and Teriaca, L. 2011, \ssr, 158, 267--288,
  \doi{10.1007/s11214-010-9698-z}.

\bibitem[Belcher\ (1971)]{belcher1971}
Belcher, J.~W. (1971), \apj, 168, 509, \doi{10.1086/151105}

\bibitem[B{\"{u}}rgi \& Geiss (1986)]{Burgi1986}
B{\"u}rgi, A. and Geiss, J. 1986, \solphys, 103, 347-383, \doi{10.1007/BF00147835}

\bibitem[Caffau et~al.\ (2011)]{Caffau2011}
Caffau, E., Ludwig, H. G.,  Steffen, M., Freytag, B., and Bonifacio, P. 2011, \solphys, 268, 255--269, \doi{10.1007/s11207-010-9541-4}

\bibitem[Cohen et al.\ (2007)]{cohen2007}
  Cohen, O., Sokolov, I.V., Roussev, I.I., et al. 2007, \apjl, 654, L163

\bibitem[Cranmer \& van Ballegooijen\ (2005)]{cranmer2005}
Cranmer, S.~R., A.~A. van Ballegooijen 2005, \apjs, 156, 265--293

\bibitem[Cranmer et al.\ (2007)]{cranmer2007}
Cranmer, S.~R., van Ballegooijen, A. A., and Edgar R. J. 2007, \apjs, 171, 520--551,
  \doi{10.1086/518001}

\bibitem[Cranmer\ (2009)]{cranmer2009}
Cranmer, S.R. 2009, Living Rev. Solar Phys., 6, 1

\bibitem[Culhane et al.\ (2007)]{culhane2007}
Culhane, J. L., Harra, L. K., James, A. M. et al. 2007, \solphys, 243, 19--61, \doi{10.1007/s01007-007-0293-1}

\bibitem[De Pontieu et al.\  (2007)]{depontieu2007}
De Pontieu, B., McIntosh, S.~W., Carlsson, M., et al. 2007, Science, 318, 1574,
  \doi{10.1126/science.1151747}

\bibitem[Dere et al.\ (1997)]{Dere1997}
Dere, K.~P., Landi, E., Mason, H. E., Monsignori Fossi, B. C., and Young, P. R. 1997,  \aaps, 125, 149--173, \doi{10.1051/aas:1997368}

\bibitem[Dere\ (2007)]{Dere2007}
Dere, K.~P.  2007, \aap, 466, 771--792, \doi{10.1051/0004-6361:20066728}

\bibitem[Dere et al.\ (2009)]{Dere2009}
Dere, K.~P., Landi, E., Young, P. R., et al. 2009, \aap, 498, 915--929, \doi{10.1051/0004-6361/200911712}

\bibitem[Evans et al.\ (2012)]{evans2012}
Evans, R.~M., Opher, M., Oran, R., et al. 2012, \apj, 756, 155, \doi{10.1088/0004-637X/756/2/155}

\bibitem[Esser \& Edgar (2000)]{Esser2000}
Esser, R., and Edgar, R. J. 2000, \apjl, 532, L71--L74, \doi{10.1086/312548}

\bibitem[Esser \& Edgar\ (2001)]{Esser2001}
	Esser, R. and Edgar, R. J. 2001, \apj, 563, 1055-1062, \doi{10.1086/323987}
 
 \bibitem[Feldman et al.\ (1992)]{feldman1992}
Feldman, U., Mandelbaum, P., Seely, J.~F., Doschek, G.~A., and Gursky, H. 1992, \apjs, 81, 387--408, \doi{10.1086/191698}

\bibitem[Feldman \& Laming\ (2000)]{Feldman2000}
	Feldman, U. and Laming, J.~M. 2000, \physscr, 61,222, \doi{10.1238/Physica.Regular.061a00222}

\bibitem[Feldman \& Widing(2003)]{Feldman2003}
Feldman, U., and Widing, K. G. 2003, \ssr, 107, 665--720, \doi{10.1023/A:1026103726147}
 
\bibitem[Feldman et al.\ (2007)]{Feldman2007}
Feldman, U., Landi, E., and Doschek, G. A. 2007, \apj, 660, 1674--1682, \doi{10.1086/513729}
   
\bibitem[Fisk et al.\ (1998)]{Fisk1998}
Fisk, L.~A., Schwadron, N.~A., and Zurbuchen, T. H. 1998,  \ssr, 86, 51--60, \doi{10.1023/A:1005015527146}
   
\bibitem[Fisk\ (2003)]{Fisk2003}
Fisk, L.~A. 2003, Journal of Geophysical Research (Space Physics), 108, 1157,
  \doi{10.1029/2002JA009284}

\bibitem[Fisk \& Zhao (2009)]{FiskZhao2009}
Fisk, L.~A., and Zhao, L. 2009, in {IAU Symposium}, vol. 257, edited by N.~{Gopalswamy} and D.~F. {Webb}, pp.
  109--120, \doi{10.1017/S1743921309029160}.

\bibitem[Frazin et~al.\ (2005)]{Frazin2005}
Frazin, R.~A., Kamalabadi, F., and Weber, M. A. 2005, \apj, 628, 1070--1080, \doi{10.1086/431295}.

\bibitem[Frazin et~al.\ (2009)]{Frazin2009}
{Frazin}, R.~A., {V{\'a}squez}, A. M. , and Kamalabadi, F. 2009, \apj, 701, 547--560, \doi{10.1088/0004-637X/701/1/547}.

\bibitem[Freeland \& Handy (1998)]{freeland1998}
Freeland, S.~L., and Handy, B. N. 1998, \solphys, 182, 497--500, \doi{10.1023/A:1005038224881}

\bibitem[Geiss et al.\ (1995)]{Geiss1995}
Geiss, J., Gloeckler, G., and von Steiger, R. 1995, \ssr, 72, 49--60, \doi{10.1007/BF00768753}.

\bibitem[Gloeckler et al.\ (1992)]{Gloeckler1992}
Gloeckler, G., Geiss, J., Balsiger, H., et al. 1992, \aaps, 92, 267--289.

\bibitem[Gloeckler et al.\ (2003)]{Gloeckler2003}
Gloeckler, G., Zurbuchen, T. H., and Geiss, J. 2003, {Journal of Geophysical Research (Space Physics)},
  108, 1158, \doi{10.1029/2002JA009286}

\bibitem[Gosling\ (1997)]{Gosling1997}
Gosling, J.~T. 1997, in {Robotic Exploration Close to the Sun: Scientific Basis},
  {American Institute of Physics Conference Series}, vol. 385, edited by
  S.~R. {Habbal}, pp. 17--24, \doi{10.1063/1.51743}.

\bibitem[Gruesbeck et al.\ (2011)]{Gruesbeck2011}
Gruesbeck, J.~R., Lepri, S. T., Zurbuchen, T. H., and Antiochos, S. K. 2011, \apj, 730, 103, \doi{10.1088/0004-637X/730/2/103}

\bibitem[Hahn et al.\ (2010)]{Hahn2010}
Hahn, M., Bryans, P., Landi, E., Miralles, M. P., and Savin, D. W. 2010, \apj, 725, 774--786, \doi{10.1088/0004-637X/725/1/774}

\bibitem[Hahn et al.\ (2012)]{Hahn2012}
Hahn, M., Landi, E., and Savin, D. W. 2012, \apj, 753, 36, \doi{10.1088/0004-637X/753/1/36}

\bibitem[Hollweg\ (1986)]{Hollweg1986}
Hollweg, J.~V. 1986, \jgr, 91, 4111--4125, \doi{10.1029/JA091iA04p04111}

\bibitem[Howard et al.\ (2008)]{howard2008}
	Howard, R.~A., Moses, J.~D., Vourlidas, A., et al. 2008, \ssr,  136, 67-115
	
\bibitem[Hundhausen et~al.\ (1968)]{Hundhausen1968}
Hundhausen, A.~J., Gilbert, H.~E., and Bame, S. J. 1968, \jgr, 73(17), 5485–5493, \doi{10.1029/JA073i017p05485}.

\bibitem[Jin et al.\ (2012)]{jin2012}
Jin, M. , Manchester, W.~B., van der Holst, B. et al. 2012, \apj, 745, 6,
\doi {10.1088/0004-637X/745/1/6}

\bibitem[Jin et al.\ (2013)]{jin2013}
	Jin, M., Manchester, W.~B., van der Holst, B. et al. 2013, \apj, 773, 50,
	 \doi{10.1088/0004-637X/773/1/50}

\bibitem[Klimchuk\ (2006)]{klimchuk2006}
Klimchuk, J.~A. (2006), \solphys, 234, 41--77, \doi{10.1007/s11207-006-0055-z}

\bibitem[Ko et~al.\ (1997)]{Ko1997}
Ko, Y.-K., Fisk, L.~A., Geiss, J., Gloeckler, G., and Guhathakurta, M. 1997, \solphys, 171, 345--361

\bibitem[Ko et al.\ (1998)]{Ko1998}
Ko, Y.-K., Geiss, J., and Gloeckler, G. 1998, \jgr, 103, 14539--14546

\bibitem[Kohl et al.\ (2006)]{kohl2006}
Kohl, J.~L., Noci, G., Cranmer, S. R., and Raymond, J. C. 2006, \aapr, 13, 31--157, \doi{10.1007/s00159-005-0026-7}

\bibitem[Kovalenko\ (1981)]{Kovalenko1981}
Kovalenko, V.~A. 1981, \solphys, 73, 383--403, \doi{10.1007/BF00151689}

\bibitem[Laming \& Lepri\ (2007)]{Laming2007}
Laming, J.~M., and Lepri, S. T. 2007, \apj, 660, 1642--1652, \doi{10.1086/513505}
  
\bibitem[Laming\ (2009)]{Laming2009}
Laming, J.~M. 2009, \apj, 695, 954--969, \doi{10.1088/0004-637X/695/2/954}

\bibitem[Laming\ (2012)]{laming2012}
Laming, J.~M. (2012), \apj, 744, 115, \doi{10.1088/0004-637X/744/2/115}

\bibitem[Landi\ (2007)]{landi2007}
	{Landi}, E. (2007),{Ion Temperatures in the Quiet Solar Corona}, \apj, 663,1363, \doi{10.1086/517910},
	
\bibitem[Landi et al.\ (2012a)]{landi2012a}
Landi, E., Gruesbeck, J. R., Lepri, S. T., and Zurbuchen, T. H 2012a, \apj, 750, 159, \doi{10.1088/0004-637X/750/2/159}.

\bibitem[Landi et al.\ (2012b)]{landi2012b}
{Landi}, E.,  Gruesbeck, J.~R., Lepri, S. T., Zurbuchen, T. H. and Fisk, L. A.  2012b, \apj, 761, 48, \doi{10.1088/0004-637X/761/1/48}.

\bibitem[Landi et al.\ (2013)]{Landi2013}
Landi, E., Young, P. R.,  Dere, K. P., Del Zanna, G., and Mason, H. E. 2013,  \apj, 763, 86, \doi{10.1088/0004-637X/763/2/86}

\bibitem[Landi et~al.\ (2014)]{Landi2014}
Landi, E., Oran, R., Lepri, S. T., et al. 2014, \apj, accepted.

\bibitem[Lepri et al.\ (2001)]{Lepri2001}
Lepri, S.~T., Zurbuchen, T. H., Fisk, L. A. et al. (2001), \jgr, 106, 29,231--29,238, \doi{10.1029/2001JA000014}

\bibitem[Lepri \& Zurbuchen\ (2004)]{Lepri2004}
Lepri, S.~T., and Zurbuchen, T. H. 2004, {Journal of Geophysical Research
  (Space Physics)}, 109, A01112, \doi{10.1029/2003JA009954}

\bibitem[Lionello et al.\ (2014a)]{Lionello2014a}
	Lionello, R., Velli, M., Downs, C. et al. 2014, \apj, 784, 120, \doi{10.1088/0004-637X/784/2/120}
  
\bibitem[Lionello et al.\ (2014b)]{Lionello2014b}
	Lionello, R., Velli, M., Downs, C., Linker, J.~A. and Miki{\'c}, Z. 2014, \apj, 796, 111,
      \doi{10.1088/0004-637X/796/2/111}

\bibitem[Matthaeus et al.\ (1999)]{matthaeus1999}
Matthaeus, W.~H., Zank, G. P., Oughton, S., Mullan, D. J., and Dmitruk, P. 1999, \apjl, 523, L93--L96, \doi{10.1086/312259}

\bibitem[McComas et~al.\ (2000)]{McComas2000}
McComas, D.~J., Barraclough, B.~L., Funsten, H.~O., et~al. 2000, \jgr, 105, 10419--10434, \doi{10.1029/1999JA000383}.
  
\bibitem[McComas et al.\ (2003)]{McComas2003}
McComas, D.~J., Elliott, H. A., Schwadron, N.~A.,  et al. 2003, \grl, 30, 1517, \doi{10.1029/2003GL017136}

\bibitem[McComas et al.\ (2008)]{McComas2008}
{McComas}, D.~J., Ebert, R. W., Elliott, H. A., et al. 2008, \grl, 35, L18103, \doi{10.1029/2008GL034896}

\bibitem[McIntosh et~al.\ (2011)]{mcintosh2011}
McIntosh, S.~W., de Pontieu, B., Carlsson, M., et al. 2011, \nat, {475}, 477--480, \doi{10.1038/nature10235}

\bibitem[Oran et~al.\ (2013)]{oran2013}
Oran, R., van der Holst, B., Landi, E., et al.  2013, \apj, 778, 176, \doi{10.1088/0004-637X/778/2/176}

\bibitem[Oran et~al.\ (2014)]{oran2014}
Oran, R.,  Landi, E., van der Holst, B., Sokolov, I. V., and Gombosi, T. I. 2014, \apj, under review, arXiv1401.0565.

\bibitem[Roussev et al.\ (2003)]{roussev2003}
  Roussev, I.I., Gombosi, T.I., Sokolov, I.V., et al. 2003, \apjl, 595, L57
  
\bibitem[Scherrer et~al.\ (1995)]{Scherrer1995}
Scherrer, P.~H., Bogart, R.~S., Bush, R.~I., et~al. 1995, \solphys, 162, 129--188, \doi{10.1007/BF00733429}

\bibitem[Schwenn \& Marsch\ (1990)]{Schwenn1990}
Schwenn, R., and Marsch, E., 1990, Physics and Chemistry in Space, 20.
  
\bibitem[Shearer et al.\ (2012)]{Shearer2012}
	Shearer, P., Frazin, R.~A., Hero, III, A.~O. and Gilbert, A.~C. 2012, \apjl, 749, L8
	
\bibitem[Sokolov et al.\ (2013)]{sokolov2013}
Sokolov, I.~V., van der Holst, B., Oran, R., et al., 2013, \apj, 764, 23, \doi{10.1088/0004-637X/764/1/23}.

\bibitem[Stakhiv et al.\ (2014)]{Stakhiv2014}
Stakhiv, M., Landi, E., Lepri, S. T., Oran, R. and Zurbuchen, T. H. (2014), \apj, under review

\bibitem[Suess\ (1979)]{Suess1979}
Suess, S.~T. 1979, \ssr, 23, 159--200, \doi{10.1007/BF00173809}

\bibitem[Suess et al.\ (2009)]{suess2009}
Suess, S.~T., Ko, Y.-K., von Steiger, R., and Moore, R. L. 2009, {Journal of Geophysical Research (Space Physics)}, {114}, A04103, \doi{10.1029/2008JA013704}

\bibitem[Suzuki\ (2006)]{suzuki2006}
  Suzuki, T.~K. 2006, \apjl, 640, L75
  
\bibitem[Suzuki et al.\ (2013)]{suzuki2013}
 Suzuki, T.~K., Imada, S., {Kataoka}, R. et al. 2013, \pasj, 65, 98

\bibitem[Usmanov et~al.\ (2000)]{Usmanov2000}
Usmanov, A.~V., Goldstein, M.~L., Besser, B. P. and Fritzer, J.~M. 2000, \jgr, 105, 12,675--12,696, \doi{10.1029/1999JA000233}

\bibitem[van der Holst et al.\ (2010)]{vanderholst2010}
van der Holst, B., Manchester, IV, W.~B.,  Frazin, R.~A., et al. 2010, \apj, 725, 1373, \doi{10.1088/0004-637X/725/1/1373}

\bibitem[van der Holst et al.\ (2014)]{vanderholst2014}
van der Holst, B., Sokolov, I. V.,  Meng, X., et al. 2014,  \apj, 782, 81, \doi{10.1088/0004-637X/782/2/81}

\bibitem[V{\'a}squez et al.\ (2010)]{Vasquez2010}
V{\'a}squez, A.~M., Frazin, R.~A., and Manchester, IV, W.~B. 2010, {\apj}, {715}, 1352, \doi{10.1088/0004-637X/715/2/1352}

\bibitem[von Steiger et~al.\ (1995)]{vonSteiger1995}
von Steiger, R., Schweingruber, R.~F.~W., Geiss, J., and Gloeckler, G. 1995, Advances in Space
  Research, 15, 3

\bibitem[von Steiger et~al.\ (2000)]{vonSteiger2000}
von Steiger, R., Schwadron, N.~A., Fisk, L. A., et al. 2000, \jgr, 105, 27,217--27,238, \doi{10.1029/1999JA000358}.

\bibitem[von Steiger et~al.\ (2001)]{vonSteiger2001}
von Steiger, R., Zurbuchen, T. H.,  Geiss, J., et al. 2001, \ssr, 97, 123--127, \doi{10.1023/A:1011886414964}

\bibitem[Wang \& Sheeley\ (1990)]{Wang1990}
Wang, Y.-M., and Sheeley, N. R. 1990,  \apj, 355, 726--732, \doi{10.1086/168805}

\bibitem[Wang et~al.\ (2000)]{Wang2000}
Wang, Y.-M., Sheeley, N. R.,  Socker, D. G., Howard, R. A. and Rich, N. B. 2000,  \jgr, 105, 25,133--25,142, \doi{10.1029/2000JA000149}.

\bibitem[Withbroe\ (1988)]{Withbroe1988}
Withbroe, G.~L. (1988), \apj, 325, 442--467, \doi{10.1086/166015}

\bibitem[Zhao et al.\ (2009)]{Zhao2009}
Zhao, L., Zurbuchen, T. H., and Fisk, L. A. 2009, \grl, {36}, L14104, \doi{10.1029/2009GL039181}

\bibitem[Zurbuchen et al.\ (1999)]{Zurbuchen1999}
Zurbuchen, T.~H., Hefti, S., Fisk, L. A., Gloeckler, G., and von
  Steiger, R. 1999, \ssr, 87, 353--356, \doi{10.1023/A:1005126718714}

\bibitem[Zurbuchen et al.\ (2001)]{Zurbuchen2001}
	Zurbuchen, T.~H. 2001,{Heliospheric Magnetic Field Configuration and its Coronal Sources in: "Recent Insights into the Physics of the Sun and Heliosphere: Highlights from SOHO and Other Space Missions"}, {IAU Symposium}, {203},585
 
\bibitem[Zurbuchen et al.\ (2002)]{Zurbuchen2002}
Zurbuchen, T.~H., Fisk, L. A., Gloeckler, G., and von Steiger, R. 2002, \grl, 29, 1352, \doi{10.1029/2001GL013946}.
  
 \bibitem[Zurbuchen \& von Steiger\ (2006)]{Zurbuchen2006}
Zurbuchen, T.~H., and von Steiger, R., 2006, {On the Solar Wind Elemental
  Composition: Constraints on the Origin of the Solar Wind}, in
  {SOHO-17. 10 Years of SOHO and Beyond}, {ESA Special
  Publication}, vol. 617.
  
  \bibitem[Zurbuchen\ (2007)]{Zurbuchen2007}
Zurbuchen, T.~H. 2007, \araa, 45, 297--338, \doi{10.1146/annurev.astro.45.010807.154030}.


\end{thebibliography}
\end{document}